\documentclass[final,5p,times,twocolumn]{elsarticle}
\usepackage{lineno,hyperref}
\usepackage{amsmath}
\usepackage{graphicx,color}
\journal{Materials Today Advances}

\usepackage{multirow}
\usepackage{numcompress}

\biboptions{sort&compress}
\newcommand\etal{{\it et\ al.\ }}

\begin{document}

\begin{frontmatter}

\title{Investigating Phase Transition and Morphology of Bi-Te Thermoelectric System}

\author[ssb]{Vishal Thakur}
\address[ssb]{Advanced Functional Materials Lab., Dr. S. S. Bhatnagar University Institute of Chemical Engineering \& Technology,
Panjab University, Chandigarh - 160 014, India}
\author[phys]{Kanika Upadhyay}
\author[ssb]{Ramanpreet Kaur}
\author[phys]{Navdeep Goyal}
\author[ssb]{Sanjeev Gautam\corref{mycorrespondingauthor}}
\cortext[mycorrespondingauthor]{Corresponding author} \ead{sgautam@pu.ac.in}
\address[phys]{Department of Physics, Panjab University, Chandigarh - 160 014, India}

\begin{abstract}
The optimization of secondary phase in thermoelectric(TE) materials can  {helps in improvisation of material's efficiency}.  Being a potential {contender} for lower temperature TE application, bismuth telluride(Bi$_2$Te$_3$) nanoparticles were synthesized via different routes and profiles to optimize their pure single phase. Systematic characterizations were performed with the help of X-ray diffraction (XRD), Rietveld refinement and field effect-scanning electron microscopy(FE-SEM) for structural and morphological behavior, while TE properties such as Seebeck coefficient, electrical conductivity and power-factor were measured for the purest sample chosen. Rietveld refinement in the XRD pattern of the samples revealed that only a small amount ($\sim$ 1.6\%) of Bi$_2$Te$_3$ was formed in co-precipitation method, while the hydrothermal technique increases this phase with increment in synthesis duration. This work focused on the phase evolution of Bi$_2$Te$_3$ with increasing synthesis duration time at constant temperature and vice-versa. XRD and Rietveld refinement revealed that the hydrothermal technique (150 $^\circ$C for 48 hours) can synthesize purest samples (84\% Bi$_2$Te$_3$ phase in this case).  {FE-SEM and Energy Dispersive X-ray analysis unveiled that the impure phases in the system {were} quantitatively reduced, and it supported by decline in  atomic percentage of oxygen from 37\% to 11\%,  in addition to this, it was also found that particle size was also decreased with increase in temperature.} { {This reduction in the particle size}, with an increase in synthesis temperature, shows a decrease in the percentage abundance of Bi$_2$Te$_3$ phase due to surface Te-desorption.} The observed electrical conductivity {of the chosen sample is} $\sim$20 times greater, while Seebeck coefficient is $\sim$3 times lower than that of pure Bi$_2$Te$_3$ phase. The detailed analysis has generalized the growth mechanism in Bi$_2$Te$_3$ phase evolution by the diffusion of Bi into Te nanorods to fabricate hexagonal Bi$_2$Te$_3$, {and Te-desorption from the surfaces of these particles}.
\end{abstract}


\end{frontmatter}

\section{Introduction and review}
The world is currently facing  {environmental hazards }and there is a great need of clean energy  \cite{r1}.  {The consumption of fossil fuels helps us to meet our energy needs but it also leads to worldwide environmental pollution, ozone depletion, global warming, etc \cite{r2}}. This encouraged researchers to  explore renewable and natural energy sources. A possible source of energy to meet the energy demand is the conversion of household or industrial waste heat  into electricity. Researchers have recently attempted to make use of heat energy generated at the garbage incinerator, and for that they have installed a 60 W module at the boiler section where temperature variation was from 823 to 973 K and an efficiency of 4.4\% was achieved with adequate cooling system \cite{r13}. The ability of thermoelectric(TE) technology to convert waste heat to electricity,  without emission of any green house {gases, by virtue of seebeck effect, makes this technology a good choice for a clean source of energy.} Furthermore, the absence of any moving parts  {in the TE module} makes this technology vibration and noise free  \cite{r4}. The  performance o {f this TE module depends} upon device design and selection of material \cite{ricn}. The efficiency of thermoelectric device is further represented in terms of figure-of-merit given as
\begin{equation}
  ZT\ = \frac{\sigma S^{2} T}{\kappa} \label{eq1}
\end{equation}
where $S$, $\kappa$ and $\sigma$  represents Seebeck coefficient, thermal conductivity and  electrical conductivity, respectively  \cite{r5}. These parameters are interdependent with  {one another} which makes manipulation of a high \emph{figure of merit} a bit difficult to achieve. {Seebeck coefficient is simply the induction of voltage in a material, when it is subjected under an temperature gradient. This induced voltage is a result of difference in the energy of Fermi levels at hot and cold ends.} Generally, $\sigma$ and $S$ vary reciprocally, so the increment of TE power factor $(\sigma S^{2})$ above the particular optimum value for bulk materials is not feasible. {Methods like `resonant states, modulation-doping, band engineering/modification, carrier pocket engineering, and energy barrier filtering' etc. are some of the techniques used to optimize the value of seebeck coefficient\cite{jia2019, paul2011dramatic, lee2010effects, liu2008enhanced}. Gayner \etal \cite{gayner17, gayner15, gayner16} have investigated the thermoelectric properties of Al, Ni, Cu doped PbSe system. { The phenomenon of `band modification' via impurity states was observed in case of Al-doped PbSe system where, Al-doping leads to a drastic increase in the effective mass and density of states, which eventually results into an increase of $S$ value\cite{gayner15}}.  {Apart from increment in effective mass, other factor such as number of charge carriers, also plays an important role in changing the $S$ value, for example, in case of Ni-doped PbSe, although, the effective mass increases with increment in Ni-doping}, but there was also an increase in the number of charge carriers, and both of these factors altering the value of $S$ in the opposite direction, thus the overall seebeck coefficient decreases with the increase in Ni-doping in PbSe\cite{gayner17}. This anomaly in seebeck coefficient values of Al- and Ni-doped samples, attributed to the position of Fermi level(E$_F$). The position of Fermi level in Al doped PbTe is rather low(E$_F$$\approx$ 17-22 meV) compared to the Ni-doped sample(E$_F$$\approx$ 28-30 meV), higher the value of E$_F$, lower will be the value of $S$, and vice versa\cite{gayner17, gayner15}. The interesting case appears in Cu-doped PbSe, where Cu-doping neither have any considerable band modification, nor introduces any resonant states, but still it affects the seebeck coefficient anonymously with change in concentration of Cu-doping\cite{gayner16}. At low concentration of Cu-doping, $S$ value increases because of the reduction in hole(charge carriers) concentration, but on further Cu-doping, the hole concentration decreases, while the concentration of electrons keeps on rising. This results {into  change in the } sign of $S$, from positive to negative,  {and such impurities which changes the type of majority of charge carriers by changing the doping concentration are known as the `Amphoteric impurities'}. A similar, amphoteric nature of Cu-doping  {was} also observed for Bi$_2$Se$_3$ systems, where at lower concentration, Cu-doped Bi$_2$Se$_3$ have holes as majority charge carriers, but as the Cu-doping increases, it {shifted to the electrons }as majority charge carriers\cite{vavsko1974amphoteric}.}

  {The `Figure of merit' of a TE material is inversely proportional to its total thermal conductivity($\kappa$), which} is a combination of thermal conductivity  {of} to the charge carrier($\kappa_e$) and phonons {($\kappa_p$)} present in the material.  {The} thermal conductivity {of} the charge carrier, is calculated from the equation $\kappa_e$ = $L_{o}$ $\sigma$ $T$, where $L_{o}$, $\sigma$ and $T$ be Lorenz number, electrical conductivity and temperature, respectively. The value of  { $\kappa_p$} can simply be calculated from the difference of total $\kappa$ and $\kappa_e$. To decrease the total thermal conductivity one can either  {decrease the $\kappa_e$ or $\kappa_p$}.
B {ut, by decreasing  $\kappa_e$ it will lead t}o reduction in $\sigma$, as $\kappa_e$ is directly proportional to the $\sigma$. So, one has to concentrate on reducing the  {$\kappa_p$} to obtain a high \emph{figure of merit} \cite{n2}. Various strategies have been adopted to control these parameters  such as phonon glass electron crystal (PGEC) \cite{r6},  doping, energy filtering, multiple band conduction mechanism and the convergence of electric bands  \cite{r5,r7}. The idea behind these processes is the reduction of  { $\kappa_p$,} either by scattering or blocking of phonons motions present in the material. Nanostructuring is a revolutionary step which made achievement of $ZT$ value up to 2 (two) possible as compare to $ZT$ $<$ 1 for bulk material, due to their enhancement of electrical and mechanical properties  \cite{r8,r9,r10,r11}. Nanostructuring also leads to the blocking of phonons, which results into the reduction of the  {$\kappa_p$} of the material. Nanotubes structures which are having low dimension and spacious cages, reduces the  { $\kappa_p$} by both blocking and strongly scattering the phonons through the cage like structures\cite{Nanotube}.
The nano thermoelectric materials with a a fine grain sizes and variety of morphologies such as nonorods, nanosheets, nanoparticles, sheet rods can be obtained by simple but effective synthesis method i.e. Solvothermal or Hydrothermal method. The variable reaction parameters such as `temperature, pressure, reaction time, solvent used, concentration of reacting species' etc. can be used to get different size, morphology, to attain the desired thermoelectric properties of the materials.

Solid solution alloying is also a process used to increase the $ZT$ of thermoelectric material. The idea behind the improvement of $ZT$ by solid solution alloying is to decrease the lattice thermal conductivity of the system, without altering its electric properties. Alloying {of the material,} gives the short range distortions, which remarkably increases the phonons scattering. Since, the charge carriers have a longer wavelength than that of phonons, therefore these short range distortions does not affects the scattering of charge carriers \cite{r1996}. {But this Solid solution alloying process has a significant drawback incomparision to hydrothermal or sovothermal process. This process is performed only at a higher temperature ,while hydrothermal or solvothermal processes can be  relatively performed at low synthesis temperatures.}

With the generation of electricity by using thermoelectric generator(TEG), other energy sources like biomass, solar, geothermal, infrared radiation have gained increased utilization in thermoelectric-based systems.Various attempts have been made to utilization TE generators in cogeneration systems to improve the overall efficiency of the system \cite{r16,r17,r18,r19}. In order to make advanced TE materials, the main challenge is to make a balance between the $ZT$ and power factor \cite{r4}. The commercial application of thermoelectric  {material} is very limited because of their low efficiency, which makes TE materials incompetent with refrigeration and conventional power generation. To compete with conventional power sources, novel thermoelectric m {aterial needs to increase its} efficiency.

Various electronically conducting polymers consisting carbon nanomaterials and nanocomposites, has also being developed, in order to achieve high $ZT$ value and lightweight thermoelectric devices  \cite{r20}.

\subsection{Bismuth Telluride as thermoelectric material}
The compound of Bismuth telluride (Bi$_2$Te$_3$) has been under intensive study due to its distinctive properties and its TE application at room temperature. Bismuth telluride is one of the simplest and commonly available low-temperature thermoelectric material. It was {first investigated for TE applications} in 1954  \cite{r21}. Both bismuth and tellurium have high atomic weight, therefore it has a high density at the grain boundaries. This results in selective diffraction of phonons by highly dense grain boundaries { as these decreases the energy of the lattice vibration or scatters those vibrations in a crystal} and eventually leads to the  {reduction of $\kappa_p$} \cite{goldsmid13}.

The crystal structure of Bi$_2$Te$_3$ consists of a hexahedral-layered structure with 5 atomic layers stacked by {\it van der wall} interaction along with the c-axis of the unit cell as shown in Fig. \ref{fig1}. Bismuth telluride  {has} a narrow band gap of 0.15 eV \cite{bg01} and low melting temperature of 585 $^\circ$C \cite{r2020}.
In Bismuth telluride, the charged carriers may be holes or electrons, which gives the $p$- type or $n$- type characteristics to it. Charged carrier concentration in Bi$_2$Te$_3$  {can} be controlled generally by the doping of elements such as Cadmium, elements of Carbon \& Nitrogen Families (Sn, Pb, As, Sb and Bi), etc. The doping of accep {tor impurities like Bi, Pb, Sb$_2$Te$_3$, Te, Se into $p$-type Bismuth telluride results in an} increases the hole concentration to the system, while the addition of compounds like halides of Cu or Ag (eg: CuBr, AgI, CuI etc.), {acts as} a donor impurities  {resulting in} the n-type Bi$_2$Te$_3$\cite{r1996}.
\subsection{Factors affecting \emph{figure-of-merit} of Bismuth Telluride}
\subsubsection{Interdependence of thermoelectric properties}
The relationship between Seebeck Coefficient and conductivity for Bi$_2$Te$_3$ is same as expected  {from} other thermoelectric materials. At the higher conductivity ranges, due to the Onset of degeneracy, the Seebeck Coefficient decreases, while at the lower conductivity range, this decrease of Seebeck coefficient is even more rapidly because of intrinsic conduction, as observed by W. M. Yim \etal \cite{r1996}. At a particular value of conductivity, the seebeck coefficient for $n$-type Bi$_2$Te$_3$ is greater than for $p$-type Bi$_2$Te$_3$, mainly in the degeneracy region \cite{r1996}.

In general, the Seebeck Coefficient depends upon the density of state and the carrier mobility.  {The carrier mobility of electrons is more than that of holes in the degeneracy region, whereas the , density of state  for the electrons is lower to that of holes}. So, the net value of Seebeck coefficient is almost same for both $n$-type and $p$-type. Even then, $p$-type Bi$_2$Te$_3$ shows a lower electrical conductivity, that is because in some cases, excess of Bi present in the system adds the degeneracy region, which decreases the mobility of holes, so the seebeck coefficient and electrical conductivity value is lesser for the $p$-type Bi$_2$Te$_3$ than that of $n$-type Bi$_2$Te$_3$  \cite{r1996}.

The thermal conductivity and electrical conductivity of a material depends directly upon the concentration of charge carriers, so, the minimization of charge carrier results into the decrease in the thermal as well as the electric conductivity. The only factor which can decrease the thermal conductivity without having any negative impact on the other parameters is the decrease in $\kappa_p$, by the process of `scattering of phonons'. Phonons are simply the lattice vibrations, and the process of coupling of these lattice vibrations with each other or with any structural defects, known as the ``Scattering of phonons". The lattice thermal conductivity of the material also alters by the presence of secondary phase. The $n$-type Bi$_2$Te$_3$ which contains halide impurities, exhibits a better coupling between halide impurity and Bi$_2$Te$_3$ lattice, which might lead to better dispersion of lattice vibrations, which in quantum terms called `more scattering of phonons', and therefore leading to smaller thermal conductivity than $p$-type Bi$_2$Te$_3$ \cite{r1996}. There are different methods to decrease the lattice thermal conductivity, and {one} such method was studied by Venkatasubramanian \etal \cite{Venkatasubramanian} and Yim \etal \cite{r1996} by studying the Bi$_2$Te$_3$/ Sb$_2$Te$_3$ superlattices which had lower lattice thermal conductivity because of more scattering of phonons in the supercrystal lattice system.

Since both the parameters Seebeck Coefficient and thermal conductivity depends upon the electrical conductivity, so electrical conductivity must be closely controlled to get high $ZT$ value.  The investigation of ZT versus electrical conductivity by Yim \etal \cite{r1996},shows that, the maximum value of $ZT$ for Bi$_2$Te$_3$ is obtained at electric conductivity nearly 1000 ohm$^{-1}$cm$^{-1}$, and showed the effect of excess of Bi and CuI-doping on TE properties of  Bi$_2$Te$_3$.
For $p$-type Bi$_2$Te$_3$ (Bi or Pb in excess) the maximum $ZT$ value is obtained at lower electrical conductivity (Z$_{max}$=2.2$\times$10$^{-3}$ deg$^{-1}$ at 900 $\Omega^{-1}$cm$^{-1}$) whereas, for $n$-type material it is obtained  at higher electrical conductivity(CuI doped Bi$_2$Te$_3$) (Z$_{max} = 2.6\times10^{-3}$ deg$^{-1}$ at 1100 $\Omega^{-1}$ cm$^{-1}$) \cite{r1996}.

\subsubsection{Morphological dependence}
The morphology (superlattices, nanowires, nanotubes etc.) of a system depends upon synthesis technique, route and reactions conditions followed \cite{r1997,r1998}. Bismuth telluride is prepared by various processes like mechanical alloying and hot pressing, powder extrusion sintering, chemical alloying, solvothermal synthesis, hydrothermal methods, wet chemical techniques and laser ablation in solution \cite{r31,zakeri2009synthesis,toprak2003chemical,stavila2013wet,r34,watanabe2004preparation}.

The morphology affects the transport property of the system and the increase in synthesis temperature may lead to agglomeration of the tiny particles and form a plate or wafer like structure. The value of $\sigma$ and $\kappa$ decreases as morphology of sample alters from small dimensional particle to higher dimensional plates \cite{r1997}.

Researchers have followed different routes to control the morphology and phase formation in Bi$_2$Te$_3$ system. Synthesis routes have been discussed in the section with main focus on hydrothermal method because of its ability to control the shape and size of synthesized materials. One of such attempts include formation of disc like structure with diameter 74.1 nm by Yokoyama \etal  using hydrothermal method \cite{r31}.
No impurity peak was observed in the system and hydrolyzed dehydroascorbic acid was proposed as an excellent oxidation resistant capping agent. On contrary to this, two groups Yang \etal \cite{r2000}
and Kim \etal \cite{r2001} initially synthesized Te nanowires and used it as reagent template for further synthesis of Bi$_2$Te$_3$ nanowires. Zhang \etal \cite{r2002}
doped La-atoms into the lattice of Bi$_2$Te$_3$ by using bismuth chloride, lanthanum chloride and Te powder as initial precursors and sodium borohydride and sodium hydroxide as reducing agent and pH balance, respectively.
Feutelais \etal   \cite{r50} studied the different phases in the $Bi-Te$ system. A new phase with Bi$_8$Te$_9$ composition was found platelet-like hexagonal crystals with parameters a = 4.41 \AA. The average distance between two consecutive atom's layer was 2 \AA. The  Bi$_8$Te$_9$ can be considered to be made up from 51 layers of atoms i.e. 27 Te and 24 Bi.

 {Rashad \etal  \cite{r34} synthesized bismuth telluride} via solvothermal process in the presence of different organic modifiers and sodium hydroxide was used as alkali modifier. Bismuth nitrate pentahydrate and ethylene glycol was used for making Bi precursor solution. Also tellurium oxide, ethylene glycol was used to synthesize Te nanopowders. Hydrazine hydrate was used as a reducing agent. No extra peaks in XRD pattern indicated the formation of pure rhombohedral of Bi$_2$Te$_3$ crystal structure (JCPDS 01-082-0358). Also it showed that crystallite size increased from 26.8 nm to 32.5 nm by the addition of NaOH.  Bi$_2$Te$_3$ nanorods of 35-70 nm in diameter  were observed in SEM images. The morphology of nanoparticles was converted to nano-circular sheets with a high homogeneity due to the presence of NaOH as it increases the alkalinity which further increased the solubility of TeO$_2$.

 {Zhao \etal  \cite{r2828} facilely synthesized }low-dimensional structured Bi$_2$Te$_3$  nanocrystals with various morphologies by using Bismuth Chloride(BiCl$_3$),  Tellurium(Te), Sodium-dodecyl-benzene-sulfonate (SDBS) or polyvinyl-pyrrolidone (PVP) as surfactants. The powder was synthesized at 373 K. NaBH$_4$ as a reductant and NaOH as a pH controller was used for the experiment. The XRD pattern was indexed using datasheet of hexagonal  Bi$_2$Te$_3$ and indicated pure phase formation of  Bi$_2$Te$_3$ nanoparticles. SEM images displayed  irregular flakes with size ranging from 60-400 nm. TEM images showed that there was  Bi$_2$Te$_3$ nanotubes of length 700 nm and diameter 45-64 nm. Also  Bi$_2$Te$_3$ needle-shaped nanowires of length 2 $\mu$m and a diameter of 50 nm were in good agreement with SEM images.


 {Gupta \etal  \cite{gupta2012synthesis} synthesized Bismuth Telluride} nanostructures at different concentration of KOH (0.5 - 1.5 M) and for different reaction timings  via refluxing method using Bismuth Chloride(BiCl$_3$), Tellurium powder, Ethylenediaminetetraacetic acid (EDTA), Potassium hydroxide (KOH), Sodium borohydrid (NaBH$_4$) in inert gas atmosphere (Ar gas). It was  {evident} from XRD pattern that sizes of nanoparticles were inversely proportional to the concentration of KOH which is in well agreement with TEM images. It was also observed that increased reaction time leads to the rod-like formation of bismuth telluride nanoparticles which is a good result.

{Zhao \etal \cite{Nanotube} has synthesized the rhombohedral Bi$_2$Te$_3$ nanotubes by hydrothermal method at 150 $^\circ$C from Bismuth chloride, Tellurium powder, EDTA, sodium borohydride and sodium hydroxide as reducing agent and pH balance respectively. These nanorods reduces the $\kappa_p$ by efficiently blocking the phonons, and gives higher ZT value as compared to the zone melted Bi$_2$Te$_3$ samples.}

The thermoelectric properties of quantum dots was inspected by Foos \etal  \cite{r39} by synthesizing  Bi$_2$Te$_3$ with particle size less than 10 nm using bismuth perchlorate oxide hydrate and sodium dioctylsulfosuccinate (AOT), trioctylphosphine (TOP) in the presence of nitrogen gas. Also deionized water, absolute ethanol, toluene and hexane were purged into N$_2$ gas. X-ray diffraction pattern showed broad peaks at 27.80 $^\circ$C which corresponds to bismuth telluride peak. The average particle size measured from TEM images was 4.5 nm. This method minimized the need for special equipment, handling and reagents.


Table \ref{tab1} discusses the pure phase formation of  Bi$_2$Te$_3$ nanoparticles using different reducing agents at different reaction temperature. Morphology and dimensions of the nanoparticles also have been mentioned for above discusses papers.

\begin{table*}[htb!]  \centering
\caption{Different reducing agents, reaction temperature, morphology and the dimensions of the nanoparticles.} \vskip 0.25cm \label{tab1}
\begin{tabular*}{1.0\textwidth}{@{\extracolsep{\fill}}|l|l|l|l|l|l|} \hline
Sample& Reducing          & Reaction          &  Particles            & Particles                      & References            \\
Code  &  agent            & Temperature       & morphology            & dimensions                     &                       \\ \hline
a.  &  Ascorbic           & 30-70 $^\circ$C   & Plates                & L = 2.0 $\mu$m                 &  \cite{r31}           \\
    &  acid               &                   &  or wires             &  D=10 nm                       &                       \\
b.  & Hydrazine hydrate   &  180$^\circ$C     &  Plate-like           & T = 18 nm                      &           \cite{fubi} \\
    & and NH$_3$.H$_2$O   &                   &                       & D = 100 - 200 nm               &                       \\
c.  & Hydrazine           &  160$^\circ$C     &    1D nanowires       &  L=3-5 $\mu$m                  & \cite{kht}            \\
    & hydrate             &                   &                       &    D=200-300 nm                &                       \\
d.  &   Sodium            &	150$^\circ$C	  &  Hexagonal flakes     & D = 100 - 300 nm               &  \cite{zeffect}       \\
    & borohydrid          &	               	  &                       &	                               &                       \\
e.  & NaH$_2$PO$_2$.H$_2$O & 150$^\circ$C     &	Irregular hexagonal   & L= 200 nm                      &      \cite{zeffect}   \\
    &                     &	                  &	 flakes               &                                &                       \\
f.  &    L-ascorbic acid  &	90$^\circ$C       &   Sheets and rods     &	L=22.43 nm                     &  \cite{neffect}       \\
g.  &    Sodium           &	   90$^\circ$C    &  Sheets and rods      &  L= 21.70 nm                   &  \cite{neffect}       \\
    & borohydrid          &	                  &  	                  &                                &                       \\
h.  & L-ascorbic acid     &   90$^\circ$C     &       Nanorods 	      & L=324 - 506 nm                 &  \cite{neffect}       \\
    &  with Trizma(as     &	                  &                       & W = 51 - 71 nm                 &                       \\
    & chelating agent)    &                   &                       &                                &                       \\
i.  & Sodium borohydride  & 363 K             & Spherical, and        & D = 76 nm                      &       \cite{neffect}  \\
    & with Trizma         &                   & Sheets                & W =  520 nm - 1.28 $\mu$m      &      \cite{neffect}   \\
    & (as chelating agent) &                  &                       & L = 1.21 $\mu$m - 2.20 $\mu$m  &                       \\
j.  & EDTA                & 180$^\circ$C      &   Flower-like         & T = 10 - 30 nm                 & \cite{fubi}           \\
    &                     &                   &  nanosphere           & D = 300 nm                     &                       \\
k.  & KBH$_4$ and         &  150$^\circ$C     &  Irregular            & L = 23 nm                      & \cite{sht}            \\
    & propane-1,3-diamine &                   &   hexagonal           &                                &                       \\
    & (as complex agent)  &                   & nanocrystals          &                                &                       \\ \hline
\end{tabular*}
\end{table*}

{ {Apart from morphology, the inhomogeneities in the sample} greatly affects the thermoelectric properties of a semiconductor. These inhomogeneities in the materials depends upon the particular growth conditions during the synthesis process of material. Homogeneity in the crystal lead to uniform carrier mobility, which gives a higher electrical conductivity. Slower the growth rate results into a greater homogeneity of the material.  As the formation growth rate of both $n$-type and $p$-type Bi$_2$Te$_3$ increases, the seebeck coefficient also increases, while total thermal conductivity and thermal conductivity due to phonons decreases. Variation in electric conductivity with the growth rate doesn't shows the similar trends for $n$-type and $p$-type Bi$_2$Te$_3$. As the growth time decreases the electric conductivity for $n$-type Bi$_2$Te$_3$ increases, while for p-type Bi$_2$Te$_3$ it decreases. The $n$-type Bi$_2$Te$_3$ is generally a single phase material, but the $p$-type Bi$_2$Te$_3$ is considered as an alloy because of formation of a discontinuous phase, which results into decrease in its electric conductivity. The exception of decrease in electric conductivity is reported by Cosgrove \etal \cite{r44}  who showed that increase in the electric conductivity of $p$-type (Bi$_2$Te$_3$)$_{50}$(Sb$_2$Te$_3$)$_{48}$(Sb$_2$Se$_3$)$_2$ with decrease in growth rate. In this work, no discontinuous phase is detected, but instead a $n$-type region in the $p$-type matrix had founded. Due to the presence of more than one phase, $p$-type Bi$_2$Te$_3$ have generated phase boundaries, which  {lead} to the scattering of phonons from the phase boundaries, results into an increases the thermal conductivity.} These is two phases exists in $p$-type, even after it is prepared under an ideal conditions. So, the thermal conductivity due to phonons is almost independent of the growth rate.

{Thus, the slower growth rate results into a higher \emph{figure of merit}.}
Annealing of the sample results into the elimination of extra phases present in the sample, and gives a more homogeneous phases, which enhances the thermoelectric properties of the sample and increases the $ZT$ value.

\subsubsection{Axial dependence of ZT}
\begin{figure}[htb!]\centering
 \includegraphics[width=0.45\textwidth]{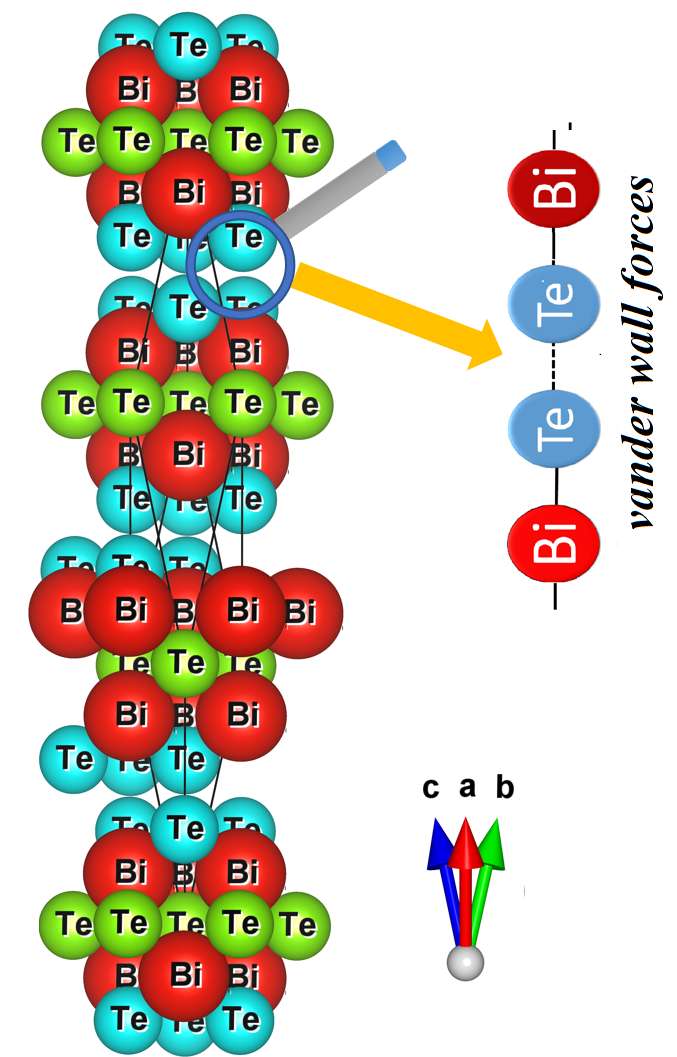}
 \caption{Bismuth Telluride structure along $c$- axis. {\it van der wall} forces exists between the adjacent Te-layers of different quintuples, whereas Bi and Te in the same quintuple are held by covalent bonds, these different types of bonding results in an anisotropic nature of Bi$_2$Te$_3$ structure\cite{goldsmid13}}\label{fig1}
 \end{figure}

Bismuth Telluride exhibits anisotropic thermoelectric properties.The atoms of Bi and Te in the crystal structure of Bi$_2$Te$_3$, are arranged in a fixed alternating sequence in the parallel layers which is repeated continuously as shown in Fig. \ref{fig1}.
Bi and Te layers are held together by covalent bond along the $c$-axis, whereas the adjacent Te-layer are held together by weak {\it van der wall} bonds. In Bismuth telluride crystals, there are two axes inclined at 60 $^\circ$ to each other which are perpendicular to the $c$-axis. The electrical and thermal conductivities are higher parallel to the cleavage planes than perpendicular to them \cite{goldsmid13}.
The electric conductivity of Bi$_2$Te$_3$ along (111) plane is 3 or 4 times greater than plane perpendicular to it, at room temperature. Thermal conductivity due to phonons along (111) plane is two times larger than the plane perpendicular to it. In exception to this, Seebeck coefficient is generally isotropic throughout the material. This results into a higher ZT value along (111) plane of Bi$_2$Te$_3$ and its alloys because of the higher charge mobility along this plane \cite{r1996}.

\subsection{Different phases of (Bi$_2$)$_m$(Bi$_2$Te$_3$)$_n$ series}
Bi-Te system exists in many intermediate phases between Bi$_2$Te$_3$ to  {bismuth bilayer(Bi$_2$)} depending upon varying concentration of Te in the compounds. These different phases formed simply by the ordered stacking of Bi$_2$ and Bi$_2$Te$_3$ building blocks. The excess of Bi$_2$ metal in a neutrally charged bilayer got stacked in between the Bi$_2$Te$_3$ blocks,  {results in} an infinitely homologous series with general formula (Bi$_2$)$_m$(Bi$_2$Te$_3$)$_n$, where ``m:n'' are the ratio of Bi$_2$ layer per Bi$_2$Te$_3$ layers. In many publications the (Bi$_2$)$_m$(Bi$_2$Te$_3$)$_n$ series structure is shown as, a hypothetical rhombohedral subcell with `abc' stacking \cite{bos2007structures,V2}. The (Bi$_2$)$_m$(Bi$_2$Te$_3$)$_n$ series can also be represented as Bi$_x$Te$_{1-x}$, where `x' is the fraction of Bi in the series. The difference in the ratio of m:n results different phases, as shown in the Table \ref{tab13} below
\begin{table}[htb!]\centering
\caption{Different phases of (Bi$_2$)$_m$(Bi$_2$Te$_3$)$_n$ series according to Bi-fraction \cite{bos2007structures,V2}.}\vskip 0.5cm
\begin{tabular*}{0.44\textwidth}{@{\extracolsep{\fill}}|l|l|c|l|}           \hline
{\bf Sr.} &    {\bf m:n}         &  {\bf Bi fraction } & {\bf Formula}   \\
{\bf No.} &                      &  {\bf (x)}          &                 \\ \hline
1.        &      0:3             &       0.40          & Bi$_2$Te$_3$    \\ \hline
2.        &      1:5             &       0.44          & Bi$_4$Te$_5$    \\ \hline
3.        &      2:7             &       0.46          & Bi$_6$Te$_7$    \\ \hline
4.        &     3:9              &       0.47          & Bi$_8$Te$_9$    \\ \hline
5.        &     1:2              &       0.50          & BiTe            \\ \hline
6.        &     3:3              &       0.57          & Bi$_4$Te$_3$    \\ \hline
7.        &     2:1              &       0.67          & Bi$_7$Te        \\ \hline
8.        &     15:6             &       0.70          & Bi$_7$Te$_3$    \\ \hline
9.        &     3:0              &       1.00          & Bi              \\ \hline
\end{tabular*}\label{tab13}
\end{table}

\begin{figure}[htb!]\centering
 \includegraphics[width=0.50\textwidth]{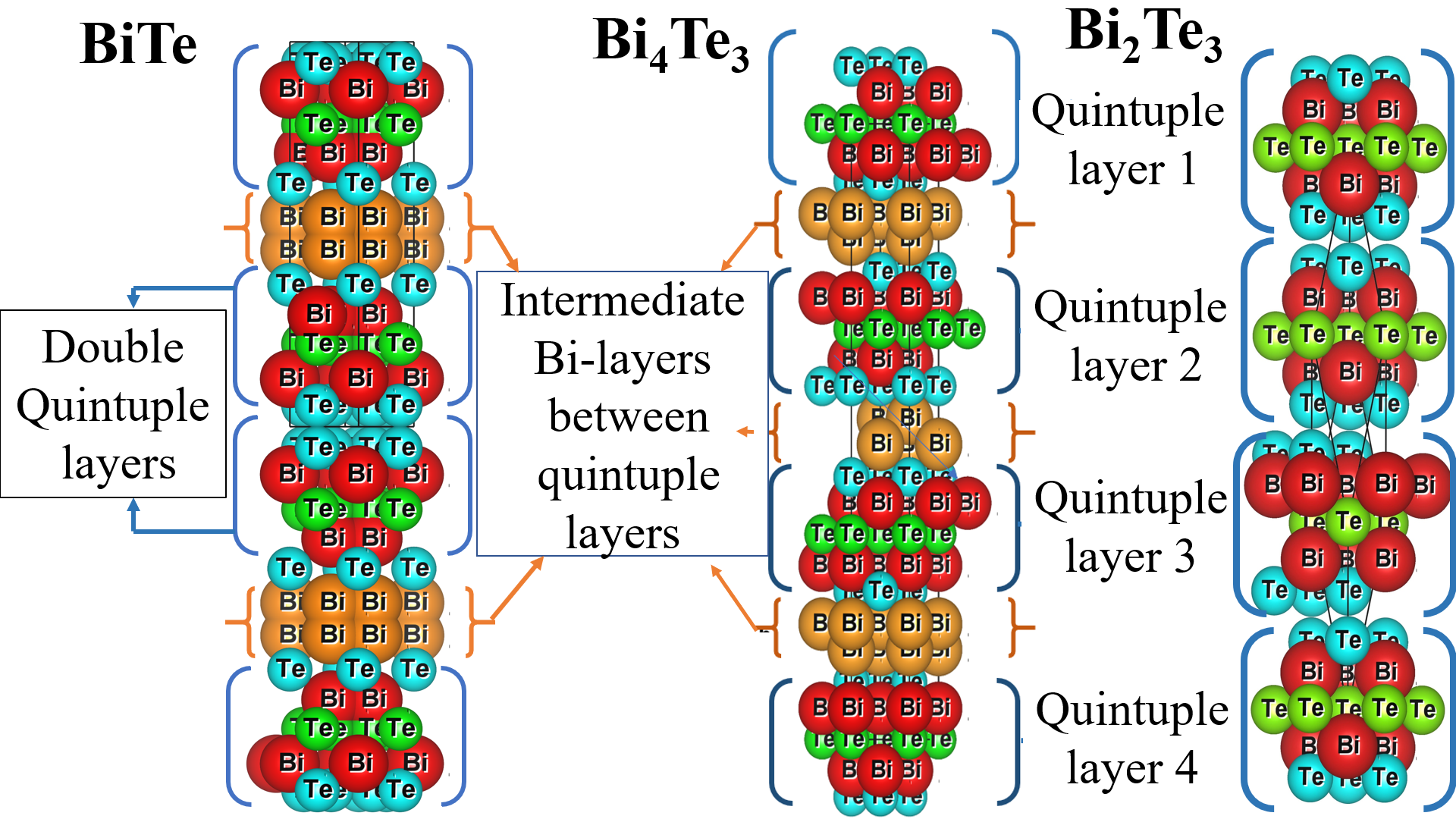}
  \caption{Schematic crystal structures of some compounds of (Bi$_2$)$_m$(Bi$_2$Te$_3$)$_n$ homologous series. As the concentration of Bi$_2$ layer increases it deposites in between the Bi$_2$Te$_3$ layers\cite{bos2007structures,V2}.}\label{fig1.5}
\end{figure}

The variation in the fraction of Bi$_2$ in different phases results into different structural and thermoelectric properties of the phases. With the variation in the composition of Bi$_2$-bilayer, the lattice constants of (Bi$_2$)$_m$(Bi$_2$Te$_3$)$_n$ series changes gradually. Bos \etal  \cite{V2} elaborated the phase change with the change in Bi fraction. In the formula Bi$_x$Te$_{1-x}$, the value of x=0.40 represents the pure Bi$_2$Te$_3$ phase, for `x' between (0.41-0.43) there exists two phases, one is of Bi$_2$Te$_3$ and the  {other} is (Bi$_2$)$_m$(Bi$_2$Te$_3$)$_n$, and as the value of `x' increases the percentage of Bi$_2$Te$_3$ phase decreases ( for x=0.41, Bi$_2$Te$_3$=83\%; x=0.42, Bi$_2$Te$_3$=73\%; x=0.43, Bi$_2$Te$_3$=43\%). While, for x=0.60-0.70, only the (Bi$_2$)$_m$(Bi$_2$Te$_3$)$_n$ phases are formed, and for 0.70 $<$ x $<$ 1.00, the elemental Bi in the form of Bi$_2$ layers, is in majority, while the Bi-Te phases of (Bi$_2$)$_m$(Bi$_2$Te$_3$)$_n$ series are limiting phases. The composition of Bi elemental phase increasing from 16\% at 0.73, to 26\% at x=0.80, to 64\% at x = 0.90.  {J. W. G. Bos \cite{bos2007structures,V2} in his work also} found that between x = 0.60 - 0.70, the XRD peaks of x = 0.60(Bi$_2$Te$_3$) and 0.63(Bi$_{63}$Te$_{37}$) are broader than for x = 0.67(Bi$_2$Te), which suggests that this broadening of peak is not simply proportional to the number of repeating units, it may be due to stacking coherence of Bi$_2$ and Bi$_2$Te$_3$ blocks. The difference in structural properties of Bi-Te phase leads to their difference in thermoelectric properties as well.

\subsection{Thermoelectric properties of (Bi$_2$)$_m$(Bi$_2$Te$_3$)$_n$ series}
The  Bi$_2$Te$_3$ pure phase is $n$-type semiconductor, which is also clear from its seebeck coefficient value (-175 $\mu$VK$^{-1}$ at 290 K). The addition of excess of elemental Bi, could even make the (Bi$_2$)$_m$(Bi$_2$Te$_3$)$_n$ series to change from $n$-type  to the $p$-type Bi-Te phase in terms of thermoelectric properties. As the fraction of Bi$_2$Te$_3$ phases decreases, resistivity of the sample increases with the increase in temperature, this shows the decrease in semiconductor behavior of the series \cite{V2}.

Seebeck coefficient depends upon the temperature as well as the fraction of elemental Bi in the (Bi$_2$)$_m$(Bi$_2$Te$_3$)$_n$ phase. The value of seebeck coefficient for the pure Bi$_2$Te$_3$ phase is -175 $\mu$VK$^{-1}$ at 290 K, while the seebeck coefficient for x = 0.44 (Bi$_4$Te$_5$) at the same temperature is -30 $\mu$VK$^{-1}$, this is due to the formation  of Bi$_2$ layer in between Bi$_2$Te$_3$ phase. At a higher Bi fraction the value of seebeck coefficient changes to be positive, behaving like $p$-type semiconductor. Generally, the Bi fraction above 0.7 always have S$>$0, in between the temperature range of 100 - 400 K except elemental Bi which have a value of seebeck coefficient of -70 $\mu$VK$^{-1}$ \cite{bos2007structures}. The  {maximum positive value of} seebeck coefficient for this series is observed for Bi$_2$Te, i.e. + 90 $\mu$VK$^{-1}$. While, some members of this homologous series like x = 0.57(Bi$_4$Te$_3$) and x=0.60(Bi$_3$Te$_2$) shows the temperature dependence of the seebeck coefficient. The value of the seebeck coefficient changes from positive to negative sign as the temperature is increased, showing the transition from $p$-type to $n$-type  thermoelectric material \cite{bos2007structures,V2}. This transition from $p$-type to $n$-type  observed at 175 K for Bi$_4$Te$_3$.

Even though the value of seebeck coefficient is changes substantially from negative value to positive value with Bi fraction, but in this change, the magnitude of seebeck coefficient is not so sufficient to get a high power factor. The maximum power factor observed in this whole series is for Bi metal i.e. 43 $\mu$WK$^{-2}$cm$^{-1}$ at 130 K, after then for Bi$_2$Te$_3$, i.e. 20 $\mu$WK$^{-2}$cm$^{-1}$ at 270 K. From the homologous series (Bi$_2$)$_m$(Bi$_2$Te$_3$)$_n$, Bi$_2$Te (m:n=5:2) and Bi$_7$Te$_3$ (m:n = 15:6) have the highest power factor of 9 $\mu$WK$^{-2}$cm$^{-1}$(at 240K) and 11 $\mu$WK$^{-2}$cm$^{-1}$(at 270K), respectively. Bos \etal  \cite{V2} had shown that the power factor of Bi$_2$Te can be quite improved by the replacement of Sb on the Bi sites. The replacement of Se on the Te sites led to Bi$_2$Te$_{0.67}$Se$_{0.33}$, has a power factor of 10 $\mu$WK$^{-2}$cm$^{-1}$ at 190 K (compared to 240 K for Bi$_2$Te).

Layered crystal structures have low thermal conductivity due to interface scattering of phonon band structure in analogy to artificial superlattice, this result into the reduction of $\kappa_p$. (Bi$_2$)$_m$(Bi$_2$Te$_3$)$_n$ series have different layered structures depending upon the Bi fraction. These layered structure results into the low value of $\kappa_p$. Sharma \etal \cite{V3} had determined the in-plane thermal conductivity due to phonons of (Bi$_2$)$_m$(Bi$_2$Te$_3$)$_n$ series and find out this homologous series had significantly lower thermal conductivity than the Bi$_2$Te$_3$ and elemental Bi.  {Sharma \etal \cite{V3}  used} the  {Debye- Callaway approximation(DCA) model for the calculation of phonons transport} and suggested that this lowered phonon's thermal conductivity is because of static defects such as point defects scattering in the crystal rather than that of unusual crystal structure. These point defects originates because of the vacancies or anti-site defects caused by the absence of Te in the crystal or excess of Bi.
Even though, there is lowering of $\kappa_p$, but this can't improve the figure of merit of the system because excess of the Bi in the series {results into decrease of seebeck \cite{V3}}.

In recent times, many experiments  {have} been performed to improve the efficiency of thermoelectric materials through the development of bismuth telluride nanostructure.

Upadhyay \etal \cite{kanika} had investigated the effect of change in the synthesis conditions on phase, morphology and electric properties of CoSb$_3$ system synthesized by solvothermal process and had successfully optimized the impure phase upto a greater extent with the increase in synthesis time. This lead us to study the effect of synthesis condition on formation of phases in Bi-Te system and morphology of  Bi$_2$Te$_3$. The phase-evaluation of co-precipitation and hydrothermally synthesized samples have also been discussed.

\section{Experimental details}
Bi$_2$Te$_3$ nanoparticles were synthesized by two different methods i.e. hydrothermal and co-precipitation using analytical grade precursors procured from Sigma Aldrich and Merck.

\subsection{Chemicals/materials}
For hydrothermal method: Bismuth chloride [BiCl$_3$], Tellurium dioxide [TeO$_2$], Hydrazine monohydrate [N$_2$H$_4$.H$_2$O], Sodium borohydride [Na(BH$_4$)], Ethanol [C$_2$H$_5$OH]. For Co-precipitation method: Bismuth nitrate pentahydrate [Bi(NO$_3$)$_3$.5H$_2$O], Nitric acid [HNO$_3$], Tellurium dioxide [TeO$_2$], Ammonia solution [NH$_3$], Hydrazine monohydrate [N$_2$H$_4$.H$_2$O], Sodium borohydride [Na(BH$_4$)] and Ethanol [C$_2$H$_5$OH]. Chemicals like  [Bi(NO$_3$)$_3$.5H$_2$O],  [TeO$_2$] and  [N$_2$H$_4$.H$_2$O] are mildly toxic, so need to be handled carefully. All chemicals were used as received during the experiment.

\subsection{Co-precipitation synthesis of Bi$_2$Te$_3$ nanocrystals}
In this method, nanocrystals of  Bi$_2$Te$_3$ were synthesized using [Bi(NO$_3$)$_3\cdot$5H$_2$O] and [TeO$_2$]. For this synthesis technique, three solutions were prepared. In the first solution, 0.1 mole of bismuth nitrate pentahydrate was dissolved in 30 ml water and 6-8 ml of HNO$_3 $ was slowly added to it as the compound is insoluble in water. Then it was left for stirring at 200 rpm for 20 minutes until a clear solution was obtained. For the second solution, 0.75 mole of TeO$_2$ was dissolved in 25 ml ammonia. 5 ml hydrazine hydrate was added to it as a reducing agent and was stirred at room temperature for 20 minutes. After that solution 1 was slowly added to solution 2 with a continuous stirring at room temperature at 250 rpm for 1 hour. To prepare the third solution, 0.1 mole of sodium borohydrid, i.e. Na(BH$_4$) was quickly dissolved in 50 ml water to obtain clear solution. Then it was transferred into a separators funnel and was slowly added to the initial mixture in the presence of N$_2$ gas. This leads to the formation of black precipitates and was stirred at 250 rpm for 1 hour. The obtained precipitates were washed three times with ethanol and three times with acetone to purify the product.The precipitates thus obtained were dried in two steps: (1). at room temperature (25 $^\circ$C) for 16 hours, (2). at 60 $^\circ$C for 24 hours in an oven. The dried precipitates were grinned in porcelain mortar to obtain fine powder.

\begin{figure}[htb!]\centering
 \includegraphics[width=0.45\textwidth]{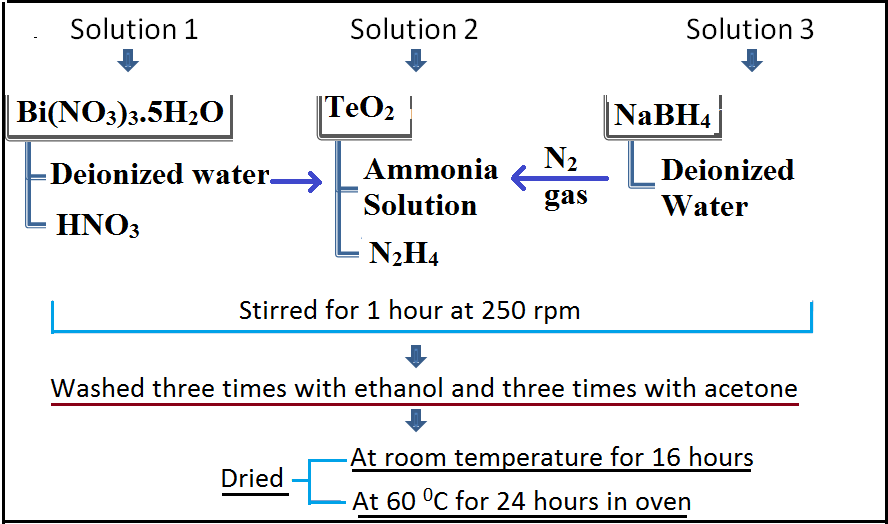}
   \caption{The schematic diagram of the recommended co-precipitation for Bi$_2$Te$_3$. }\label{fig2}
 \end{figure}

The next step was to remove additional phases by reduction with hydrazine hydrate. For this,  2 grams of the above obtained sample was transferred into the flask and were heated in the presence of N$_2$ gas. 30 ml of hydrazine hydrate was transferred into another flask and both flasks were connected to each other. Hydrazine temperature was 80 $^\circ$C and powder temperature was 300 $^\circ$C. After this, the sample was reduced to 1.8 grams. The schematic drawing of the recommended co-precipitation process is shown in Fig. \ref{fig2}. The obtained sample was further annealed in N$_2$ atmosphere at 300 $^\circ$C. The as prepared sample and annealed samples were labeled as S1 and S4 respectively.

\subsection{Hydrothermal synthesis of  Bi$_2$Te$_3$ nanoparticles}
\begin{figure}[htb!]\centering
\includegraphics[width=0.45\textwidth]{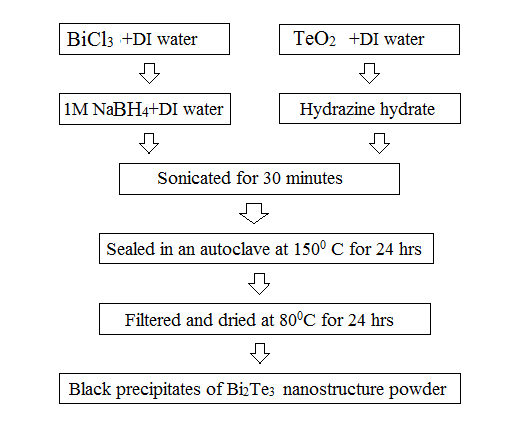}
  \caption{The schematic diagram of the improvised hydrothermal process used in the synthesis.}\label{fig3}
  \end{figure}

This route for producing  Bi$_2$Te$_3$ nanocrystals of pure phase using materials BiCl$_3$ and TeO$_2$  is simple, convenient and cost effective. In this method, 0.1 mol of BiCl$_3$ and 0.15 mol of TeO$_2$ were dissolved in 20 ml deionized water. Then NaBH$_4$  in deionized water was slowly added to Bi precursor solution. Also N$_2$H$_4$ was added to Te precursor solution. Both solutions were mixed and stirred for 10 minutes. It was then transferred to the Teflon container and sonicated for half an hour. Furthermore, it was sealed in autoclave at 150 $^\circ$C for 24 hours. It was allowed to cool down naturally at room temperature. Then it was filtered using filter paper to remove NaCl impurity and transferred into the crucible. The crucible was kept in oven for drying at 80 $^\circ$C for 24 hours. The obtained sample was crushed using pastel and mortar and was stored in the storage tube and named as H1. Similarly, four more samples were prepared, two by changing the synthesis time from 24 hours to 36 hours and 48 hours and other two by changing synthesis temperature from 150 $^\circ$C to 180 $^\circ$C and 200 $^\circ$C for 48 hours. The schematic drawing of the recommended hydrothermal process is shown in Fig. \ref{fig3}. The samples were labeled as H1, H2 and H3 for 24, 36 and 48 hours and H4 and H5 at 180 $^\circ$C and 200 $^\circ$C for 48 hours, respectively hydrothermally synthesized.

\subsection{Characterization technique}
X-ray diffraction (XRD) on the samples were performed using Rigaku equipped with Cu K$_\alpha$ radiation (1.5404$^\circ$ \AA) from 10-80$^\circ$ with a rotation speed of 5$^\circ$/min. Rietveld refinement  technique was performed using X'pert HighScore Plus to overcome the constraint of overlapping reflections. The morphology and elemental analysis on the sample were accomplished employing field emission scanning electron microscope (FESEM) and energy dispersive spectroscopy (EDS) using HITACHI, SU8010. Thermoelectric properties like Seebeck coefficient, electric conductivity and resistivity and power factor is being analyzed on Ulvac ZEM-3 at CSIR - National Physical Laboratory, New Delhi, India.  The XRD and FESEM data is analyzed using HighScore 3.0.5(Rietveld refinement)\cite{degen2014highscore} and ImageJ(FE-SEM)\cite{imagej2}, respectively.

\section{Results and discussion}
\subsection{Structural properties}
\subsubsection{X-ray diffraction analysis}
\begin{figure}[htb!]\centering
\includegraphics[width=0.48\textwidth]{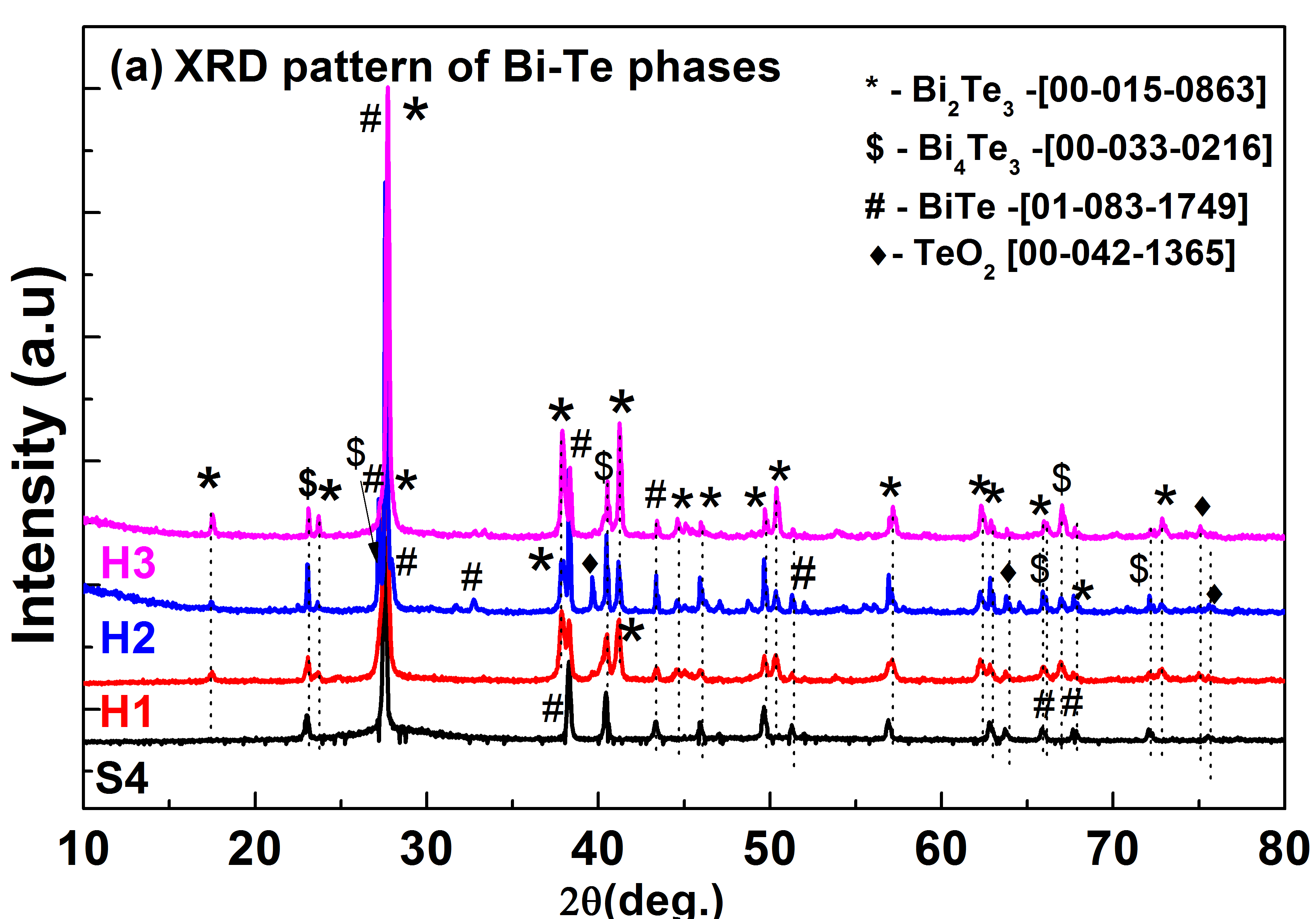}\\[0.2cm]
\includegraphics[width=0.48\textwidth]{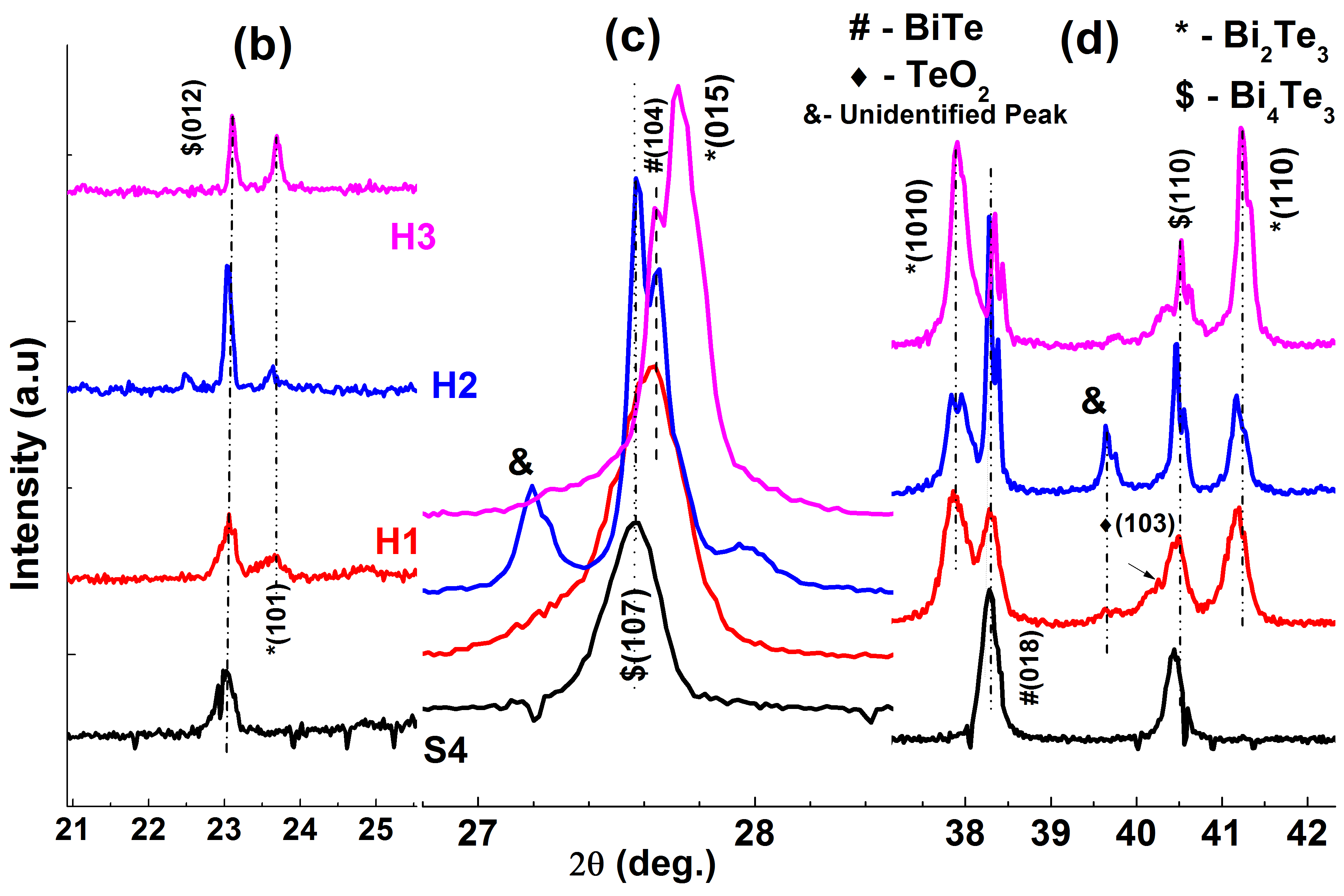}
\caption{(a) X-ray diffraction patterns of the samples prepared through co-precipitation and hydrothermal method representing different phases and (b)-(d) enlarged view of some planes evolving through various profile.}\label{fig4}
\end{figure}

Figure \ref{fig4}(a) represents the X-ray diffraction (XRD) pattern of samples synthesized through different methods and profiles. The XRD of S1 shows no peaks corresponding to Bi$_2$Te$_3$. This might be because the reaction has not attained the optimum conditions required for the formation of Bi$_2$Te$_3$ phase. On the other hand, S4, H1, H2 and H3 reveals peaks related to Bi$_2$Te$_3$ along with peaks of other impurity phases. The phases present in the sample are Bi$_2$Te$_3$, BiTe, Bi$_4$Te$_3$ and TeO$_2$ represented by $\ast$, \#, \$ and $\blacklozenge$ respectively. Some additional unidentified peaks are also  {observed} in the sample. The indexing of Bi$_2$Te$_3$  was performed with reference code 00-015-0863, BiTe by [01-083-1749], Bi$_4$Te$_3$ by [00-033-0216] and TeO$_2$ by [00-042-1365].

As in the enlarged image of XRD from Fig. \ref{fig4} (b to d),  {displays} the major Bi$_2$Te$_3$  {planes (1 0 1)}, (0 1 5), (1 0 10), (1 1 0), (3 1 1), (4 1 2), (6 5 4). Among  {these}, the highest intensity peak of (0 1 5) plane is present in H1, H2 and H3 samples at 2$\theta$ = 27.66$^\circ$, indicates the presence of Bi$_2$Te$_3$ phase. This peak is absent in S1 and S4 sample which are prepared by co-precipitation method. Although, the peak of (1 0 1), (1 0 10) and (1 1 0) plane of Bi$_2$Te$_3$ phase at 2$\theta$ =  23.68$^\circ$, 2$\theta$ = 37.83$^\circ$  and ,2$\theta$ = 41.15$^\circ$, respectively are missing in S4 sample and their intensities are increases from H1 to H3. This implies, Bi$_2$Te$_3$ phase is formed maximum in case of H3 sample. The impurity peaks of Bi$_4$Te$_3$, are  {mainly} formed at 2$\theta$ values of 23.46$^\circ$, 27.52$^\circ$, 40.49$^\circ$, representing (0 1 2), (1 0 7) and (1 1 0) planes, respectively. The intensity of the peak representing (0 1 2) plane of Bi$_4$Te$_3$ increasing from S4 to H2 system, and than decreases for H3 sample. Similarly, the peak of  (1 0 7) plane of Bi$_4$Te$_3$ at 27.52$^\circ$ is more intense for H2, then S4 system, and is absent in H1 and H3 samples. For H3 system, the other intense peaks of Bi$_4$Te$_3$ (1 1 0) and (1 0 28) planes are at 40.49$^\circ$ and 67.06$^\circ$. The peaks of Bi$_4$Te$_3$ impurity phase is most intense for H2 sample among all the samples prepared.
Bi$_4$Te$_3$ structure consists of Bi$_2$  blocks stacked  {in between} Bi$_2$Te$_3$ blocks and due to the presence of excess of Bi in the form of Bi$_2$ blocks along with Bi$_2$Te$_3$, this makes it a $p$-type material \cite{bos2007structures}.
%

The BiTe phase is another impurity found in the XRD data of the samples. BiTe impurity peaks are found at 27.59$^\circ$, 32.31$^\circ$, 38.02$^\circ$, 43.58$^\circ$, 51.42$^\circ$, 65.98$^\circ$ and 67.61$^\circ$ 2$\theta$ values representing (1 0 4), (1 0 6), (0 1 8), (1 1 4), (1 0 12), (0 0 17) planes, respectively. Among all these different phases, the peak representing (1 0 4) plane is most intense peak and is found for H1, H2 and H3 samples. At 2$\theta$ = 38.02$^\circ$, the (0 1 8) plane of BiTe is present in all the samples with H2 having most intense peak.
At 32.31$^\circ$ only the H2 sample is having BiTe peak.

This impurity is mostly found in S4 samples with impurity peaks at 2$\theta$ = 65.98$^\circ$ and 67.61$^\circ$.

Apart Bi-Te phases, some of the TeO$_2$ has also been found possibly  {formed} during the drying stage. The XRD peaks at 39.26$^\circ$, 64.10$^\circ$,74.82$^\circ$ and 75.45$^\circ$ are representing TeO$_2$ phases. These peaks are generally of low intensity. The only peak of TeO$_2$ phases at 2$\theta$ = 39.26$^\circ$ is having appreciable intensity and represents the (201) plane of TeO$_2$ phase.

\begin{figure}[htb!]\centering
\includegraphics[width=0.35\textwidth]{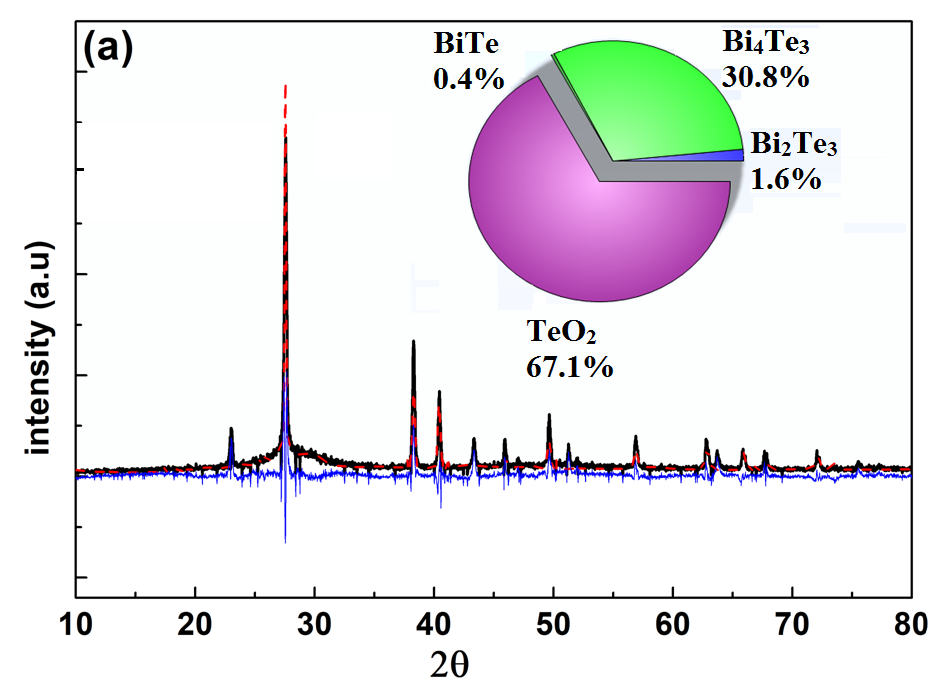}
\includegraphics[width=0.35\textwidth]{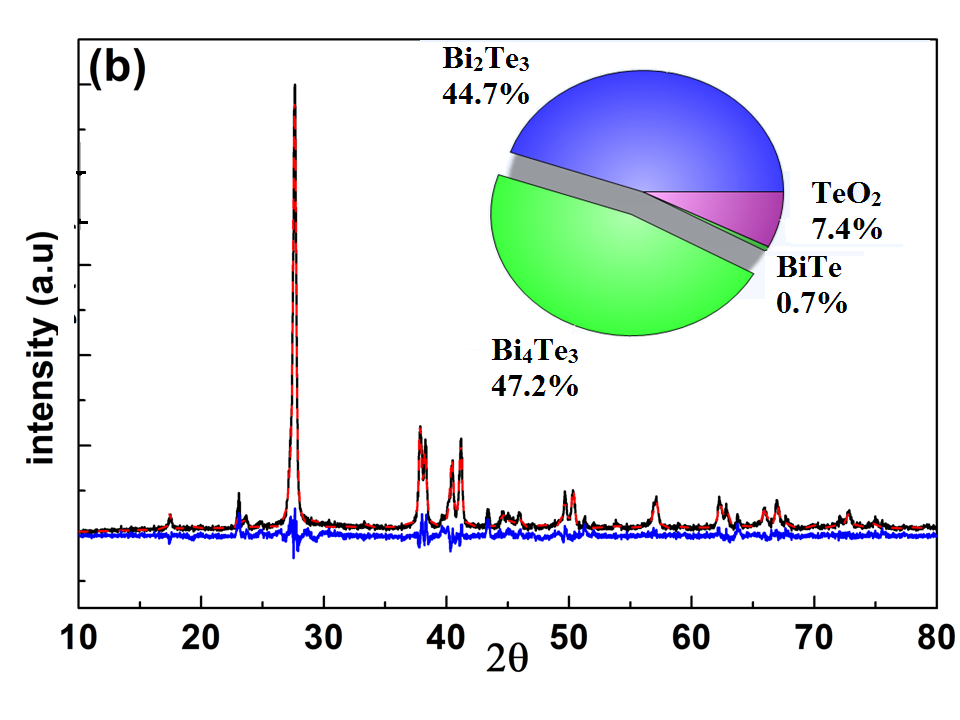}
\includegraphics[width=0.35\textwidth]{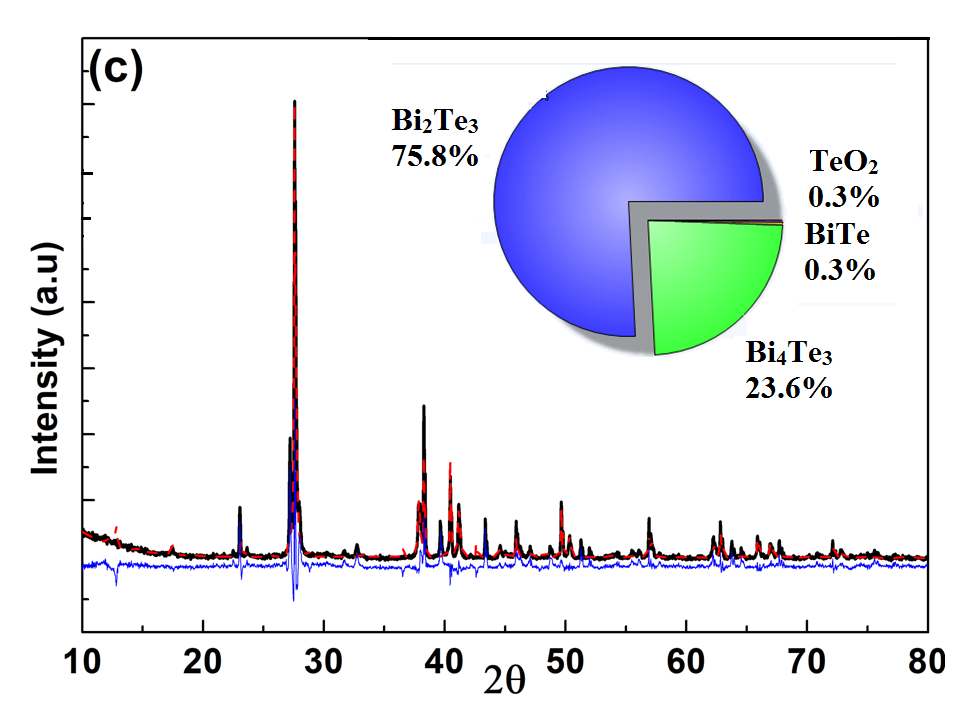}
\includegraphics[width=0.35\textwidth]{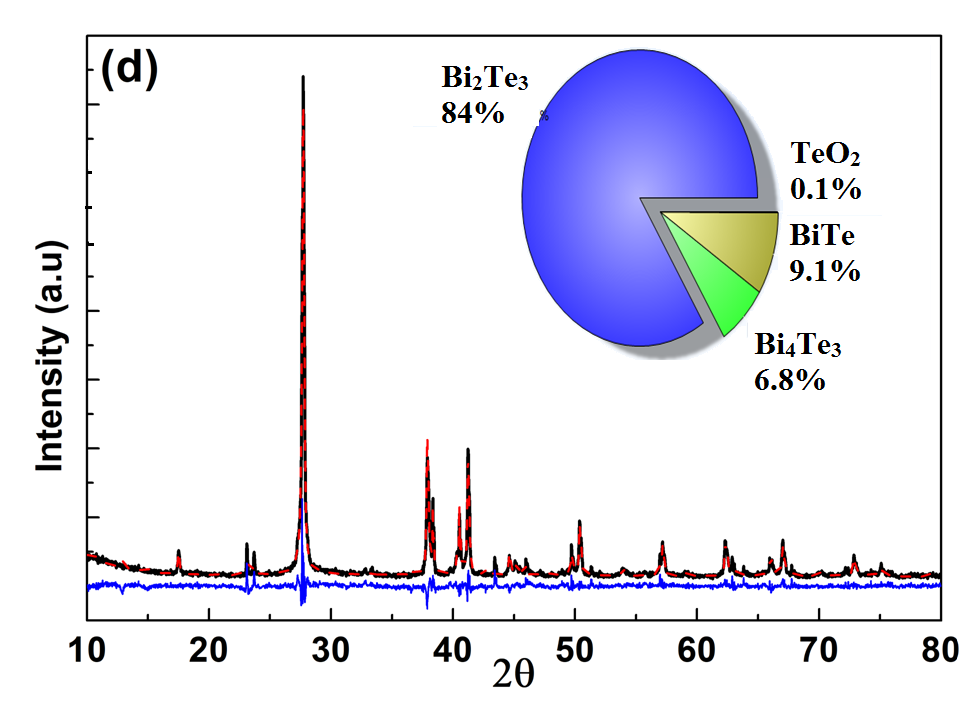}
\caption{Rietveld refinement plots (a)S4; (b)H1; (c)H2 and (d)H3}\label{fig5}
\end{figure}

\begin{figure}[htb!]\centering
\includegraphics[width=0.45\textwidth]{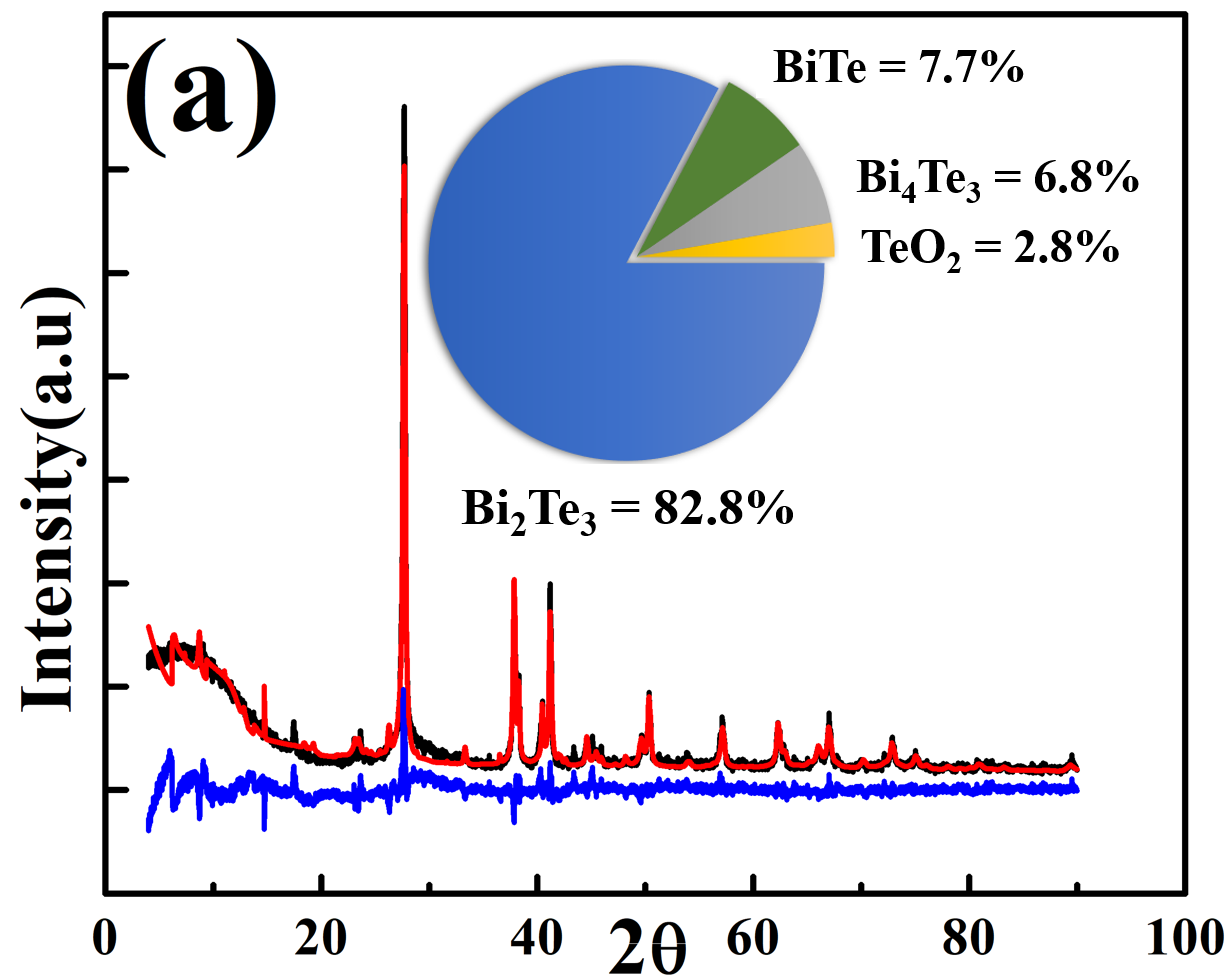}
\includegraphics[width=0.45\textwidth]{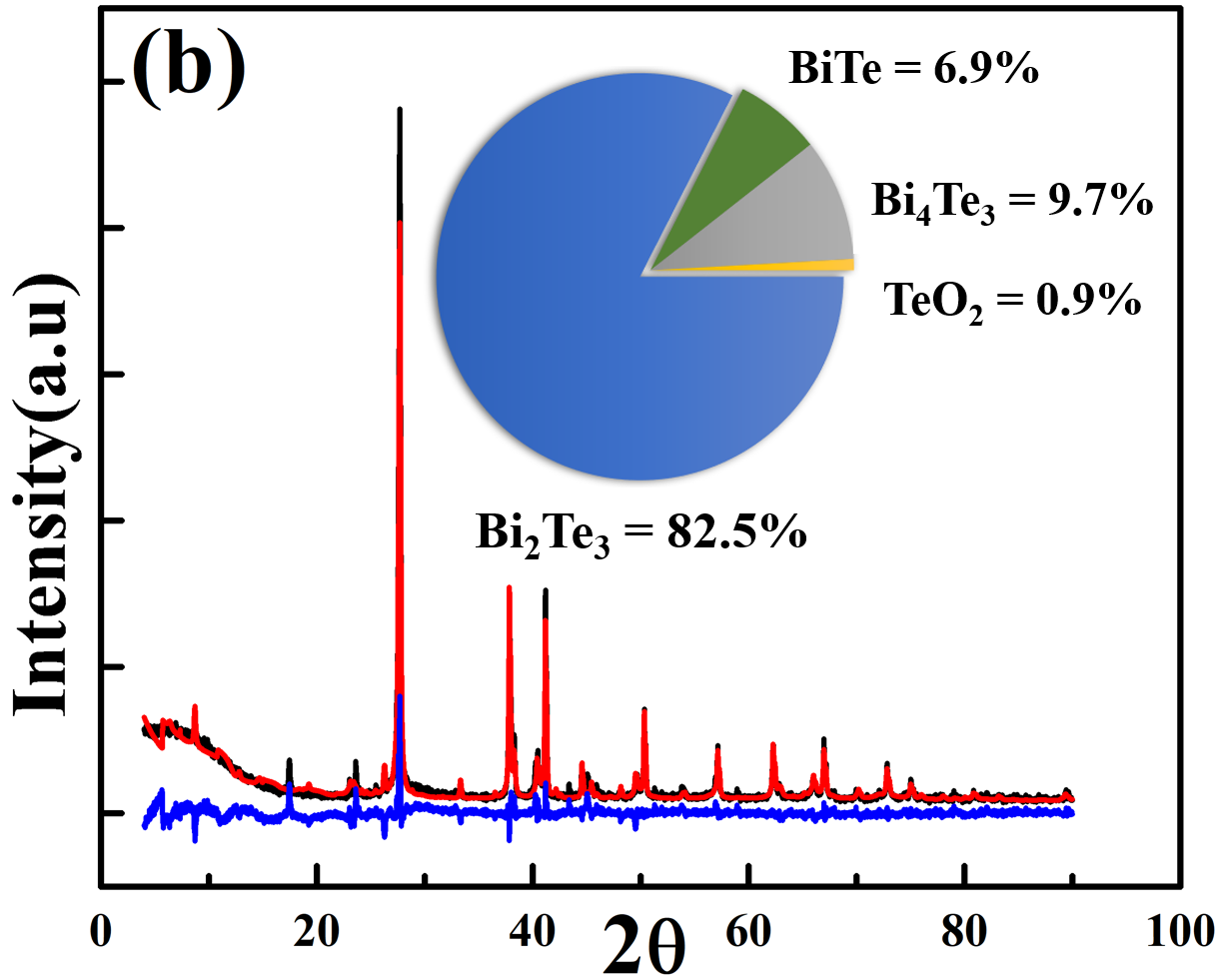}
\caption{Rietveld refinement plots (a) H4; and (b) H5}\label{fig5c}
\end{figure}

To get a better picture of the amount of distant phases present, we performed Rietveld refinement on the synthesized samples. {Fig. \ref{fig5} represents} the Rietveld refinement of the samples. It can be seen from Fig. \ref{fig5}(a) pie chart, that only 1.6\% of Bi$_2$Te$_3$ was formed with 67\% of TeO$_2$ as impurity. On contrary to co-precipitation synthesized sample, the hydrothermally synthesized samples H1, H2 and H3 showed an drastic increase in the amount of Bi$_2$Te$_3$ phase  {as} 44.7\%, 75.8\% to 84\% respectively. The amount of TeO$_2$ {decreased} 7.4\%, 0.3\% to 0.1\% for H1, H2 and H3 respectively, can be referred form Figs. \ref{fig4} (b to d). { There is a significant increase in the amount of Bi$_2$Te$_3$ phase, and other phases like Bi$_4$Te$_3$, BiTe and  TeO$_2$  decreased to a greater extent, this shows that the reaction is slow and it needs time to proceed. }From the above discussions, we noticed that, increase in synthesis time at 150 $^\circ$C results in increase of Bi$_2$Te$_3$, now let's see the effect of increase in synthesis temperature on the phase concentrations. {For this two samples synthesized} at 180 $^\circ$C (H4) \& 200 $^\circ$C (H5) has been investigated  { and pie chart in the Fig. \ref{fig5c}} represents \% of different phases present in all the samples.  The amount of Bi$_4$Te$_3$ phase remains almost same in H4 \& H5 samples, as  {that of H3 s}ample, while there is a little decrease in the \% of Bi$_2$Te$_3$, {somewhat} increase in TeO$_2$ phase, when the synthesis temperature of the samples increased. It means, at higher temperature, there occurs a loss of Te content  {in the samples}. Klimovskikh \etal \cite{klimovskikh} had also observed thermal desorption of Te atoms.  There may be a possibility of oxidation of Te to TeO$_2$ phase.  {The important point that arises from here is, the Te which is oxidized to TeO$_2$, comes out of Bi$_2$Te$_3$ phase, as it contains maximum Te:Bi atomic ratio, while other phases remains almost same(lower Te:Bi atomic ratio), as lesser the Te:Bi atomic ratio higher the thermal stability of that phase \cite{klimovskikh,bos2007structures,lind}.}

It can be concluded from the Table \ref{tab3} that by switching the synthesis route and with  {increasing} the synthesis duration, the lattic {e parameters are} decreased and the amount of Bi$_2$Te$_3$ is increased. Also, the variation in TeO$_2$ phase indicates that increasing the synthesis duration might have decreased the amount Te impurity, which eventually led to the formation of TeO$_2$ during the drying process. Thus, we {can} predict that if we further increases the synthesis duration  {and using a suitable capping agents to stop the Te desorption}, pure phase of Bi$_2$Te$_3$ sample can be attained. Table \ref{tab2} and \ref{tab3} lists the refinement parameters, refined cell parameters and amount percentage of different phases.
\begin{table}[htb!]\begin{center}
\caption{R-factor obtained for different samples during Rietveld refinement.}\vskip 0.25cm
\begin{tabular*}{0.48\textwidth}{@{\extracolsep{\fill}}|l|l|l|l|l|l|l|} \hline
\textbf{R-factor}  &{S4}  &{H1}   &{H2}   & {H3} &  {H4}  &  {H5}  \\ \hline
R$_{profile}$      &24.08 & 12.8  & 17.51 & 9.24 & 10.16  & 10.461 \\ \hline
R$_{w.profile}$    &37.87 & 17.27 & 24.08 & 12.19& 13.24  & 14.02  \\ \hline
R$_{expected}$     &12.43 & 10.94 & 9.52  & 8.84 & 6.739  & 7.139  \\ \hline
GOF                &9.29  & 2.5   & 6.39  & 1.9  & 3.86   & 3.8611 \\ \hline
\end{tabular*}\label{tab2}\end{center}
\end{table}


Table \ref{tab2} reveals that the refinement parameters and GOF for different samples were satisfactory in regards to refinement criteria.

\begin{table*}[htb!]\begin{center}\centering
\caption{Value of lattice parameter and percentage of each phase obtained from Rietveld refinement.}\vskip 0.25cm
\begin{tabular*}{0.42\textwidth}{@{\extracolsep{\fill}}|l|l|l|l|l|l|l|l|} \cline{1-8}
\multicolumn{2}{|c|}{Sample/phase} & S4     &  H1    & H2     & H3      & H4      &    H5 \\ \cline{1-8}
\multirow{3}{*}{Bi$_2$Te$_3$}  & a & 4.3867 & 4.3836 & 3.383  & 4.382   & 4.3825  & 4.446  \\ \cline{2-8}
                               & c & 30.491 & 30.462 & 30.457 & 30.474  & 30.471  & 30.48   \\ \cline{2-8}
                               &\% & 1.6    & 44.7   & 75.8   & 84.0    &   82.8  &  82.5   \\ \cline{1-8}
\multirow{3}{*}{Bi$_4$Te$_3$}  & a & 4.4590 & 4.4574 & 4.455  & 4.454   & 4.4567  & 4.4604   \\ \cline{2-8}
                               & c & 41.40  & 41.496 & 41.465 & 41.458  & 41.47   & 41.46    \\ \cline{2-8}
                               &\% & 30.8   & 47.2   & 23.6   & 6.8     & 6.8     & 6.9     \\ \cline{1-8}
\multirow{3}{*}{BiTe}          & a & 4.423  & 4.4222 & 4.4252 & 4.458   & 4.459   & 4.446    \\ \cline{2-8}
                               & c & 24.02  & 23.980 & 24.07  & 23.81   & 24.053  & 24.12    \\ \cline{2-8}
                               &\% & 0.4    & 0.7    & 0.3    & 9.1     & 7.7     & 9.7       \\ \cline{1-8}
\multirow{3}{*}{TeO$_2$}       & a & 5.08   & 4.918  & 5.504  & 8.26    & 6.88    & 7.05    \\ \cline{2-8}
                               & b & 8.38   & 8.777  & 11.77  & 4.87    & 6.07    & 5.24     \\ \cline{2-8}
                               & c & 4.19   & 4.295  & 5.585  & 3.38    & 3.50    & 3.33     \\ \cline{2-8}
                               &\% & 67.1   & 7.4    & 0.3    & 0.1     & 2.8     & 0.9      \\ \cline{1-8}
\end{tabular*}\label{tab3}\end{center}
\end{table*}

\subsection{Morphological properties}
\subsubsection{FE-SEM \& EDX analysis}
\begin{figure}[htb!]\centering
\includegraphics[width=0.23\textwidth]{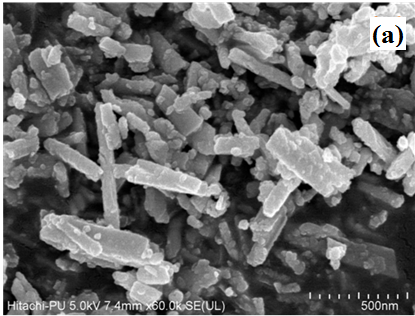}
\includegraphics[width=0.23\textwidth]{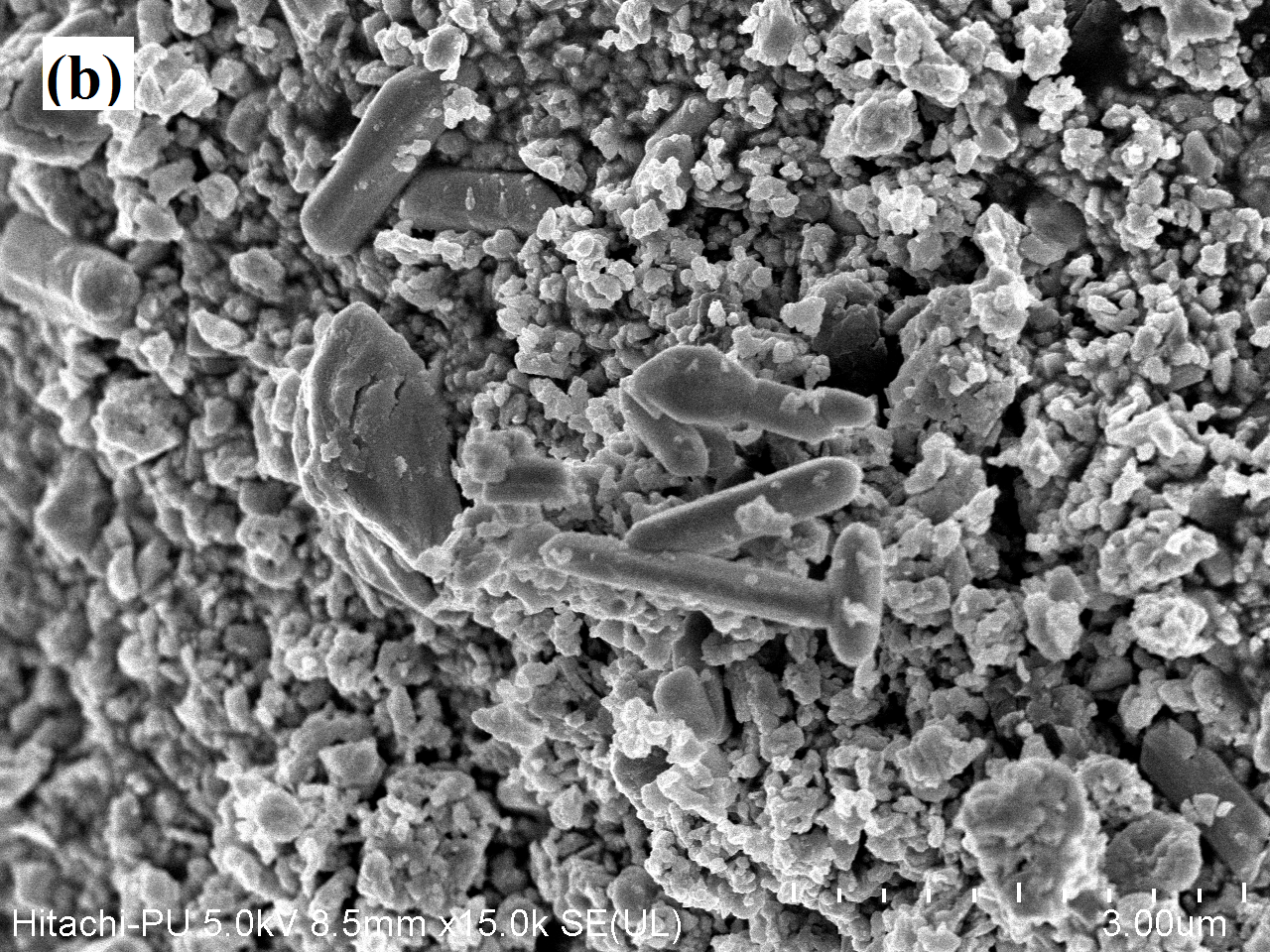}
\includegraphics[width=0.23\textwidth]{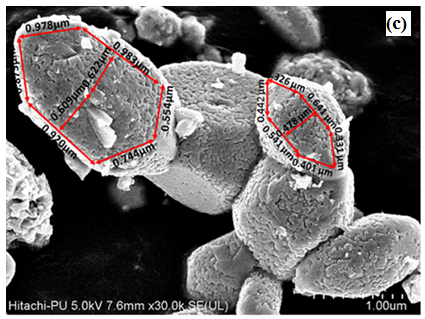}
\includegraphics[width=0.23\textwidth]{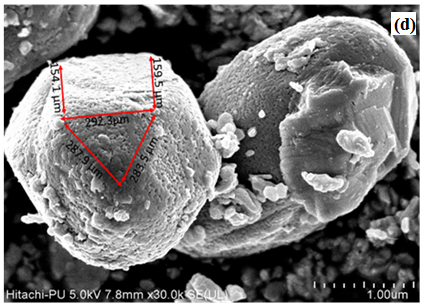}
\includegraphics[width=0.23\textwidth]{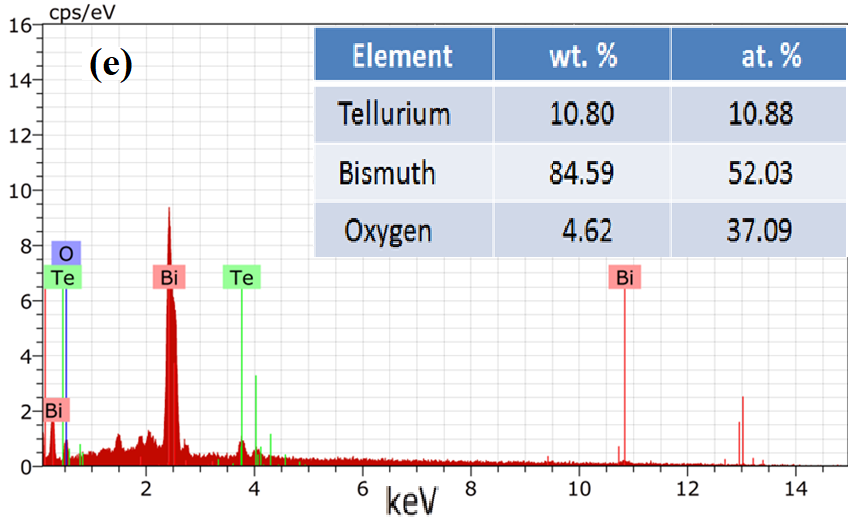}
\includegraphics[width=0.23\textwidth]{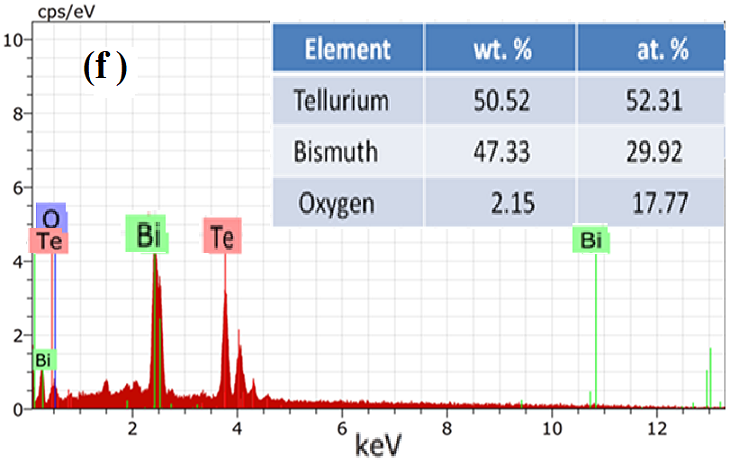}
\includegraphics[width=0.23\textwidth]{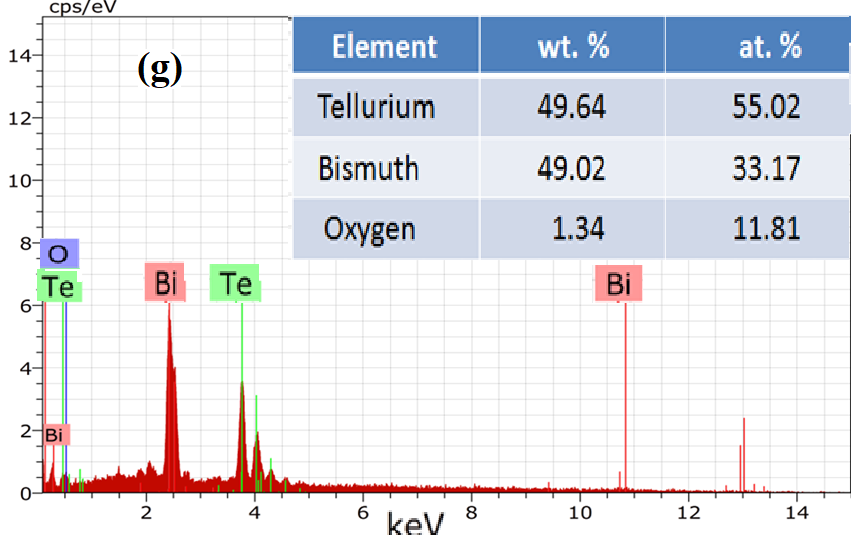}
\includegraphics[width=0.23\textwidth]{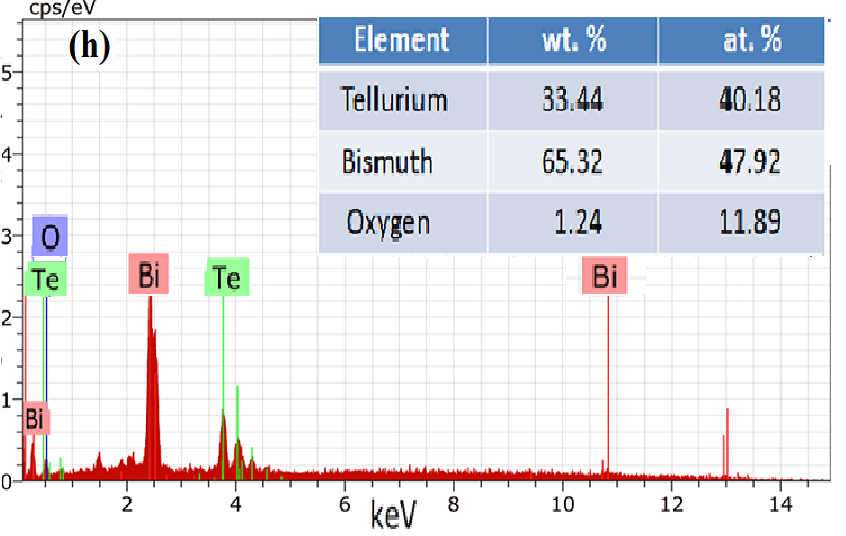}
\caption{FE-SEM and EDX image of (a \& e)S4; (b \& f)H1; (c \& g)H2 and (d \& h)H3}\label{fig6}
\end{figure}
\begin{figure*}[htb!]\centering
\includegraphics[width=0.40\textwidth]{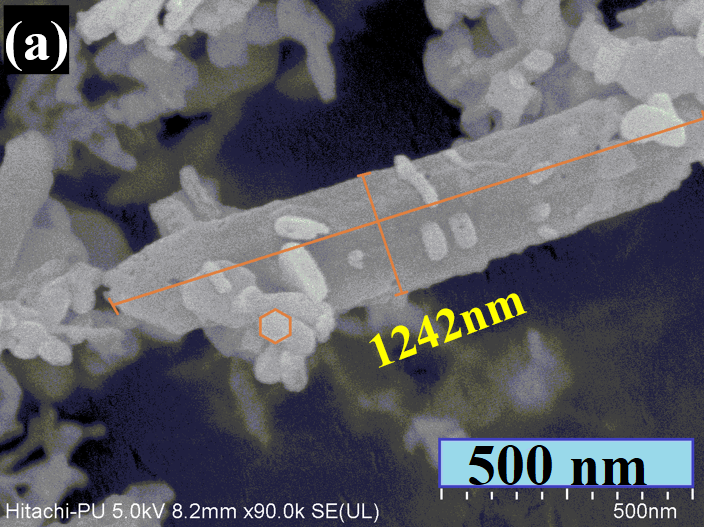}\
\includegraphics[width=0.40\textwidth]{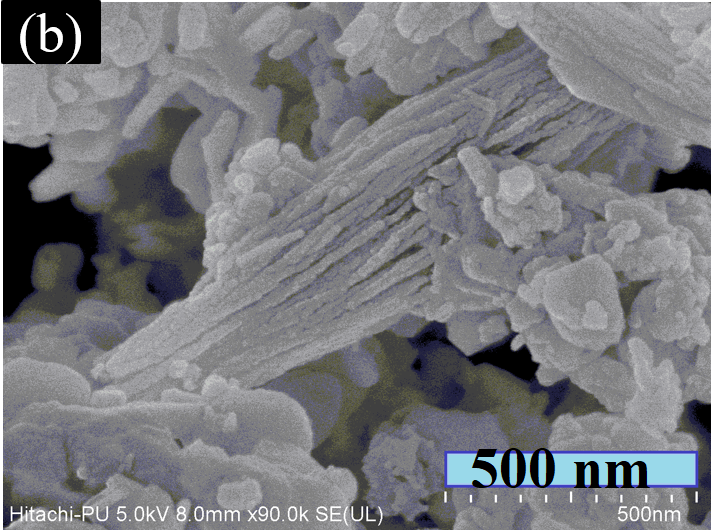}\\
\includegraphics[width=0.40\textwidth]{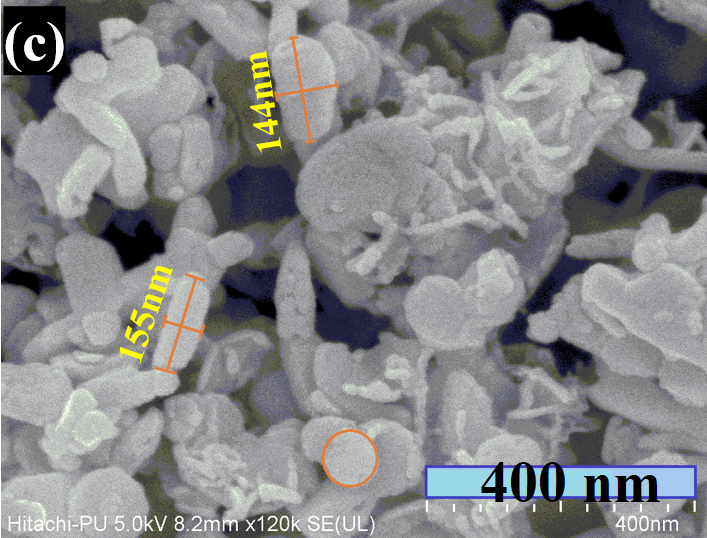}\
\includegraphics[width=0.40\textwidth]{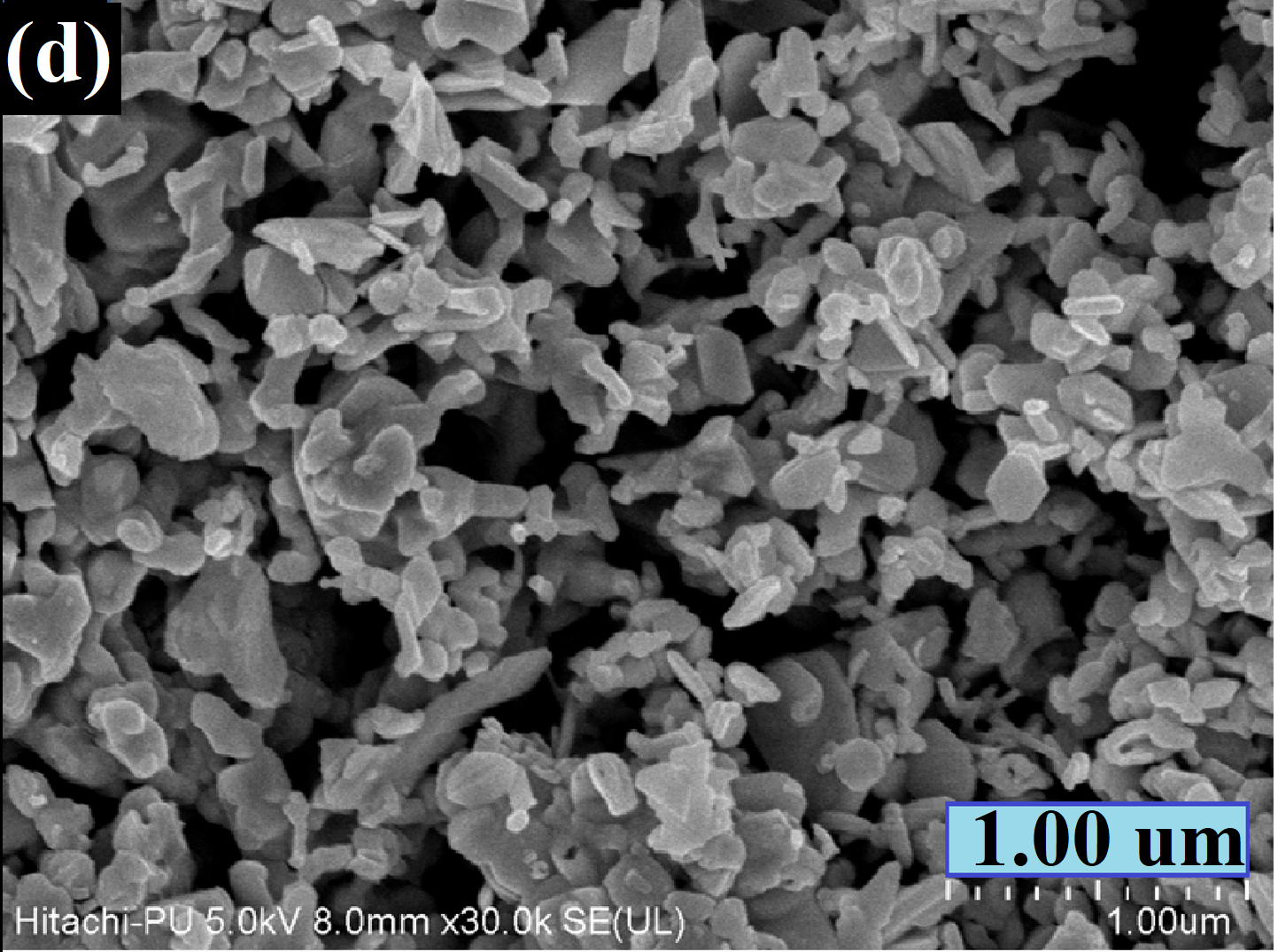}\\
\includegraphics[width=0.40\textwidth]{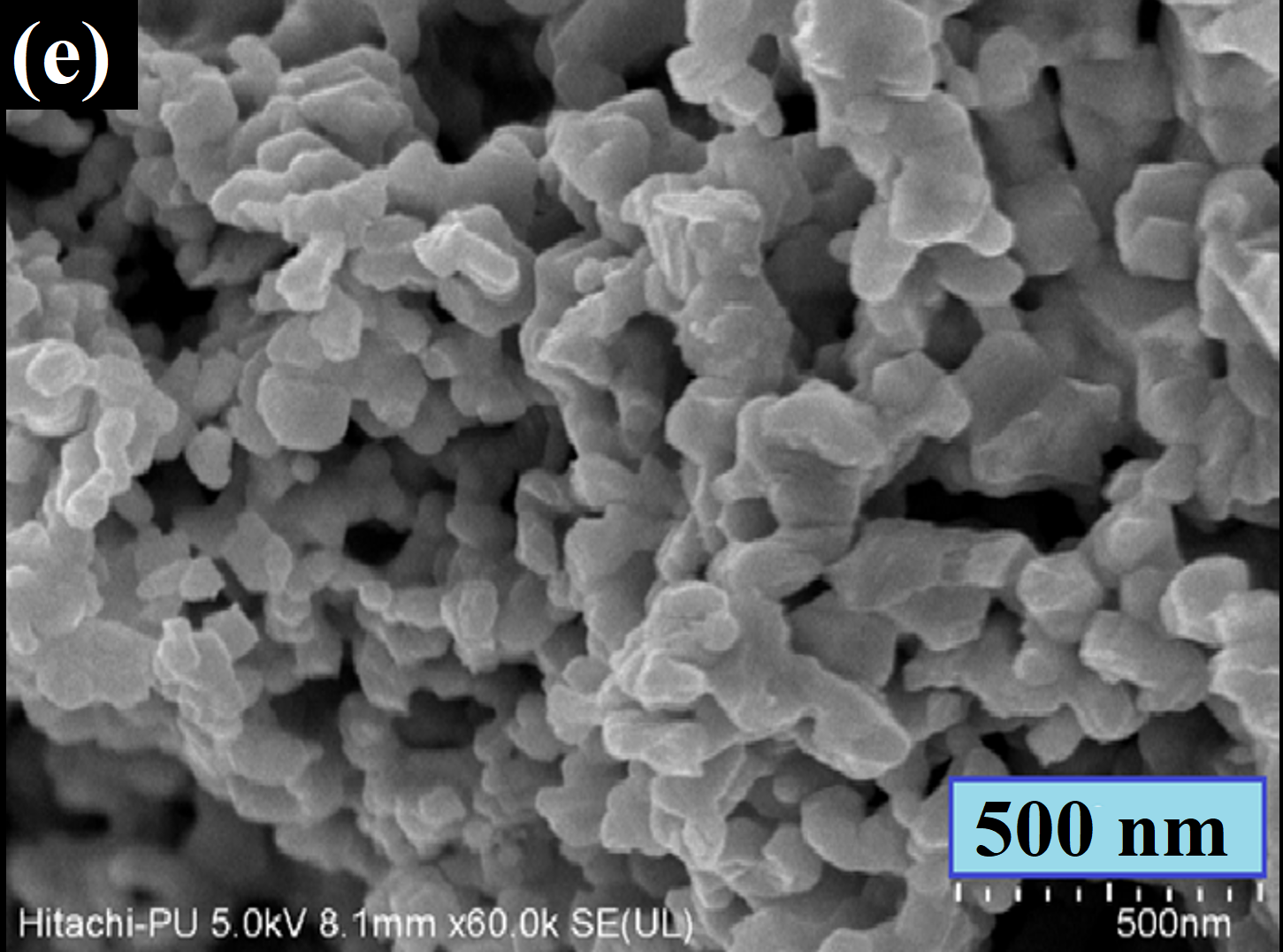}\
\includegraphics[width=0.40\textwidth]{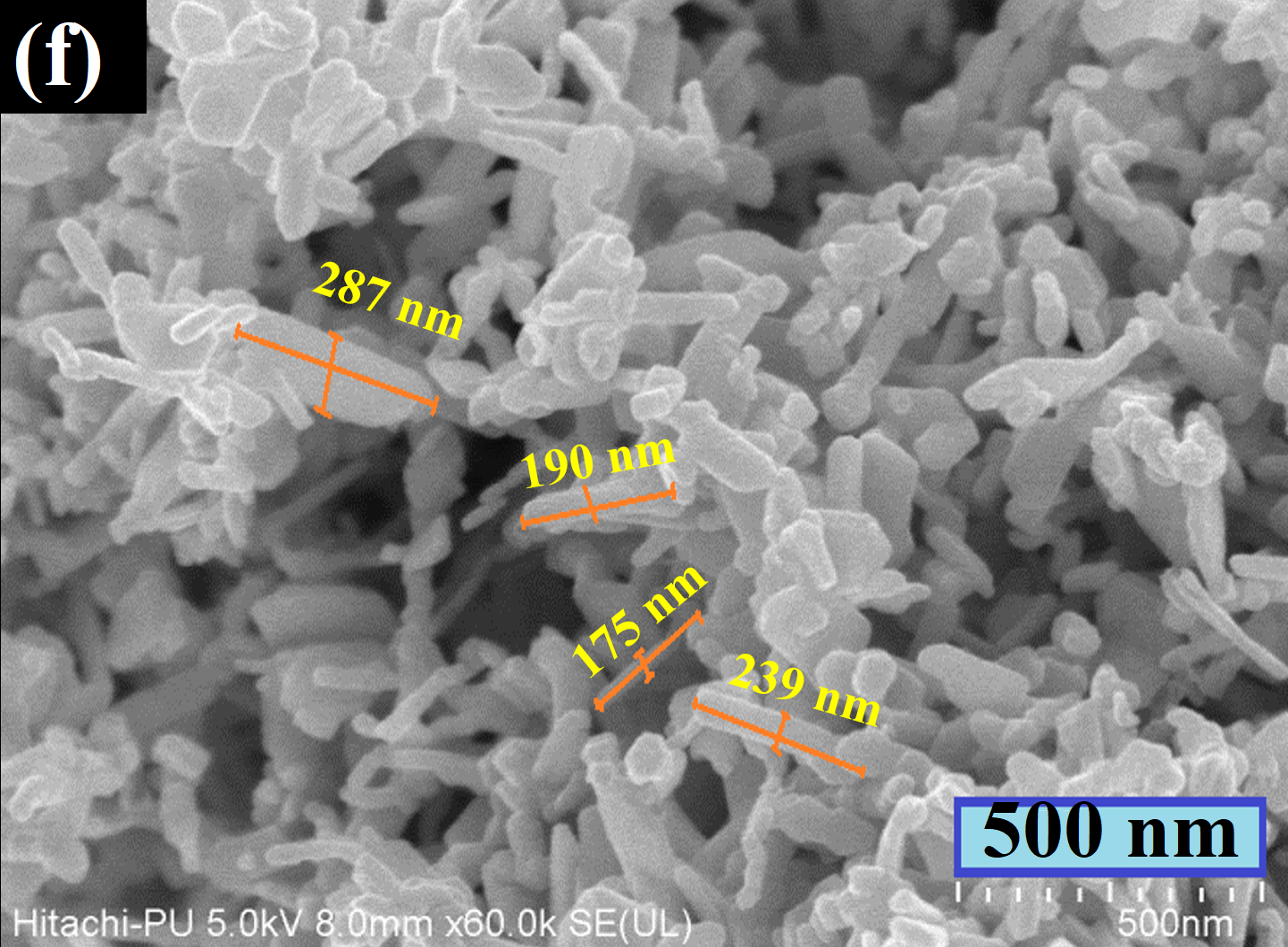}\\
\caption{FE-SEM images of (a, b \& c)H4; and (d, e \& f)H5}\label{fig7}
\end{figure*}
Figure \ref{fig6} shows the FESEM micrographs and EDX of the prepared samples.  {Figure \ref{fig6}(a) reveals }that most of the particles are formed in 200 to 250 nm range. The nanorods of TeO$_2$ was found to be 663$\times$138 nm, calculated from ImageJ software. Similarly, H1 showed presence of the nanorods with $0.9\times 1.9\mu$m dimensions. This is due to the formation of TeO$_2$ during drying stage as predicted in the previous section, no such rods were observed in H2 and H3 micrographs as the amount of TeO$_2$ was only 0.3\% and 0.1\% respectively. {Niasari \etal \cite{sht} had also observed similar results of breaking down of rod-like structure to an agglomerated form when synthesis time is increased from 12 hours to 48 hours with hydrate hydrazine a reductant.} It can be seen in the micrographs o {f H2 and H3 samples, consist o}f irregular hexagonal nanocrystals also that the bigger flakes become larger and thicker with the increase in reaction time while smaller flakes disappears. This can be explained by Gibbs-Thompson effect, the small crystals reduce in size, ions and atoms combine to form a large crystal with low free energy to form a stable system. The dimensions of the hexagonal crystals are  {labelled} in the figures.The EDX of the samples supports the observation from XRD, Rietveld refinement and FE-SEM about the formation of impure phase. It can be seen from Figs. \ref{fig6} (e),(f), (g) and (h) that, the amount of oxygen decreased from 37\% (in S4 sample) to 11\% (in H3 sample).

The FE-SEM images of H4 \& H5 samples are represented in Figure \ref{fig7}, where micrographs (a, b \& c) represents H4 sample and micrographs (d, e \& f) shows the H5 sample. There is no appreciable change in the EDX micrograph for the H4 \& H5 samples as compared to the H3 sample. Compared these images to the H3 sample, we observed that, with the increase in temperature the particle size of the sample decreases. In H3 sample, the particles are of size $\textgreater$1$\mu$m, while for H4 sample, although some particle are still of size $\textgreater$1$\mu$m as represented in Fig. \ref{fig7}(a), but most of the particles have size range $\sim$ (200-300) nm. Figure \ref{fig7}(b) clearly shows the disintegration of a micro meter rod into smaller nanorods. For H5 sample, there is no particle observed of size range in micrometers. The average size of the particle in the H5 sample ranges between (200-300)nm. This decrease in size beneficial for decreasing  {the $\kappa_p$} as a result of {scattering of phonons from grain boundaries}, which results in higher $ZT$ value.

{Their exists a great relation between the morphology of the sample and the \% abundance of different phases of the (Bi$_2$)$_m$(Bi$_2$Te$_3$)$_n$ series. Klimovskikh \etal \cite{klimovskikh} observed that the thermal desorption of the Te occurs from the surface of Bi$_2$Te$_3$ phase, and in our samples, we observed the same. In the H3, H4 \& H5 samples, the  amount of Bi$_2$Te$_3$ phase is 84.0, 82.8 \& 82.5 \%, respectively, and size of particles  {decrease} from micrometer to nanometer, as the synthesis temperature increases from 150 $^\circ$C(H3 sample) to 200 $^\circ$C(H5 sample). It may implies that, the particles of H5 sample has greater the surface by volume ratio, compared to the particles present in H3 sample, and since Te-desorption takes place from the surface of the particles, therefore, H5 samples have greater chances of Te-desorption and formation of other impurity phases, like, BiTe, Bi$_4$Te$_3$ and other phases with lower Te:Bi atomic ratio. This explained the decrease of \% of Bi$_2$Te$_3$ phase, as the synthesis temperature is increased.}

In our experiment, when hydrazine is added to the mixed solution of Tellurium dioxide and de-ionized water, under the condition that the thermodynamic energy is appropriately supplied, homogeneous Te nanorods can be obtained through the following reaction process:
\begin{equation}
TeO_2 + 2N_2H_2\cdot H_2O \rightarrow  Te   + 4H_2O + 2N_2
\end{equation}
The net reaction involved in this process can be presented as follows:

\noindent
\textbf{Co-precipitation}
\begin{equation}
Bi(NO_3)_3 + NaBH_4 \rightarrow  Bi^{3+} + NaNO_3     + BH_3 + H_2
\end{equation}
\noindent
\textbf{Hydrothermal}
\begin{eqnarray}
BiCl_3 + H_2O &\rightarrow&   BiOCl + 2HCl \label{eq4}\\
3NaBH_4 + 4HCl + 8BiOCl &\longrightarrow& \\
\rightarrow 8Bi   + 8H_2O + 4HCl &+& 3NaCl +3BCl_3  \label{eq5} \\
2Bi + 3Te &\rightarrow&        Bi_2Te_3  \label{eq6} \\
\end{eqnarray}
BiCl$_3$ selected as the Bi source is immediately hydrolyzed to form  BiOCl and HCl when distilled wa {ter is added to it. In the transformation process fro}m Te nanorods to Bi$_2$Te$_3$ nanorods, Bi$^{3+}$ ions are reduced to Bi atoms by NaBH$_4$ and then diffused into the lattice of Te nanorods to generate hexagonal Bi$_2$Te$_3$.
The study to obtain a pure phase TE material is important as the presence of impurity phases may introduce two type of types of conduction (i.e. electrons and holes). This will eventually lead to uncertainty in $\kappa_e$ (electronic thermal conductivity) introduced by bipolar effect and also the seebeck voltage developed across the ends will decrease \cite{r28,r2020}.

\subsection{Thermoelectric properties analysis}
\begin{figure}[htb!]\centering
\includegraphics[width=0.45\textwidth]{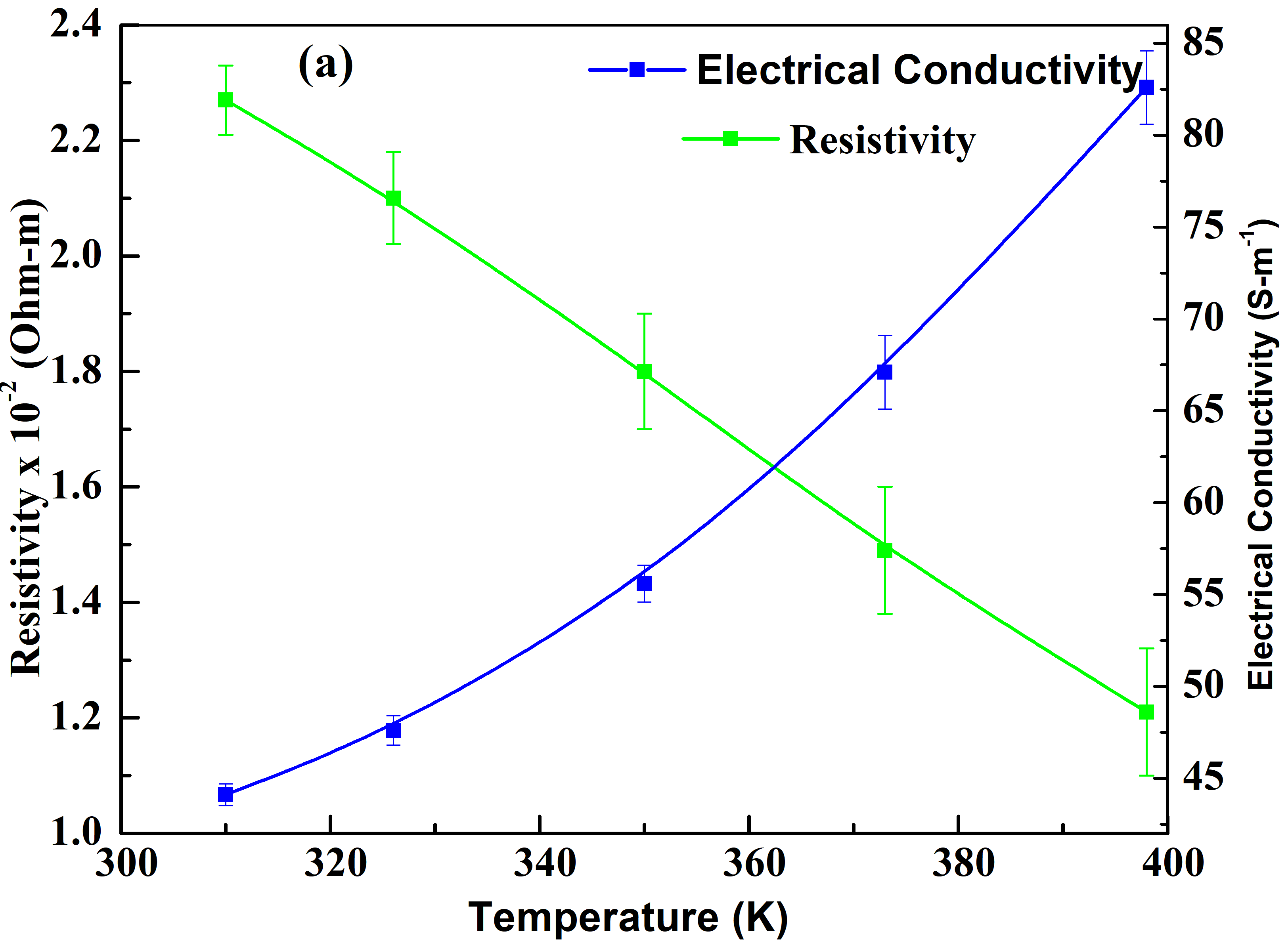}
\includegraphics[width=0.48\textwidth]{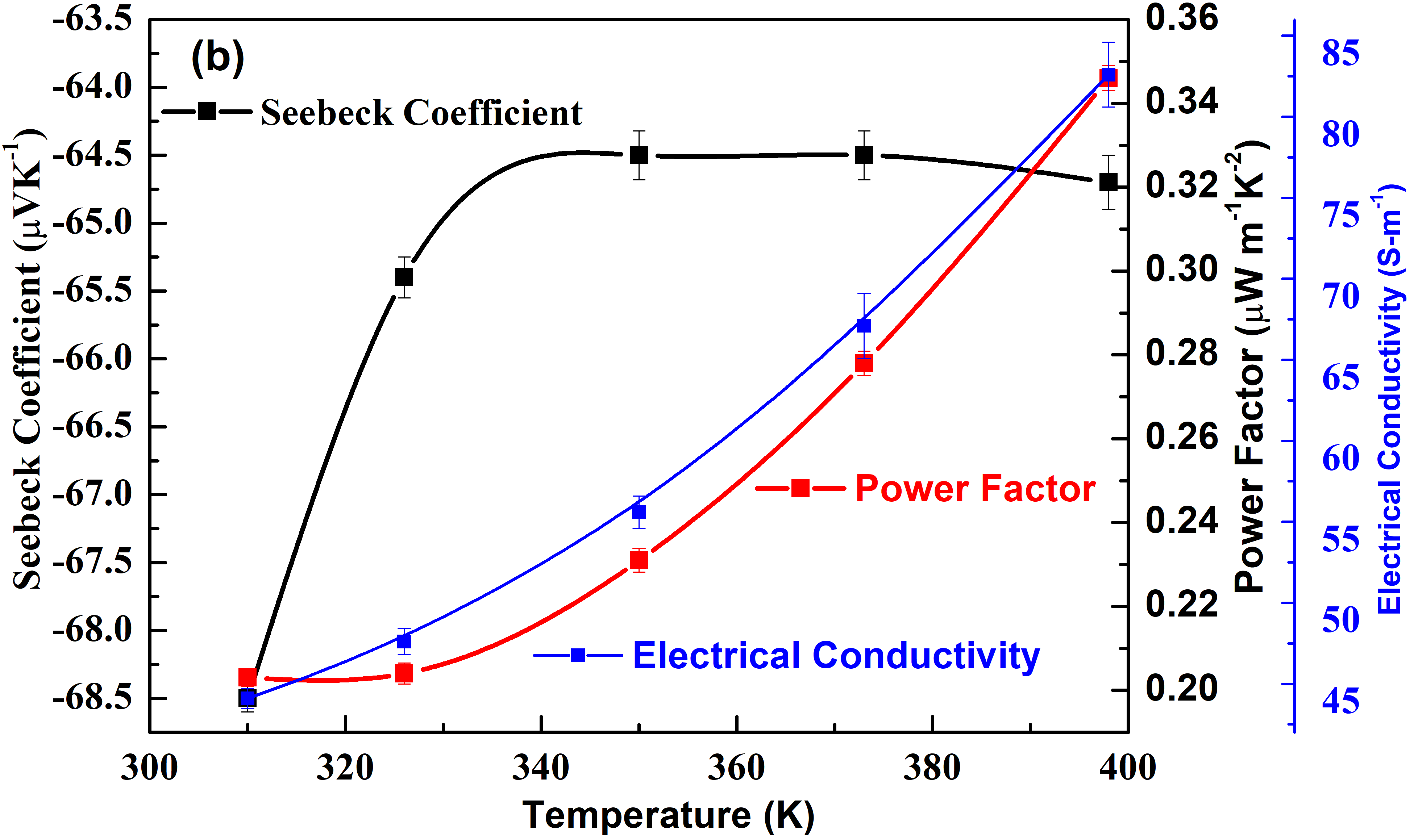}
\caption{Variation of TE properties with temperature for Bi-Te sample prepared at 200 $^\circ$C for 48 hours hydrothermally, representing, Seebeck coefficient verse Temperature(Black curve), Electrical conductivity verse Temperature(Blue curve), Electrical resistivity verse Temperature(green curve) and Power Factor verse Temperature(Red curve).} \label{fig8}
\end{figure}

Among the all samples Bi-Te prepared,the sample prepared at 200 $^\circ$C temperature for 48 hours, is the best sample to study the TE properties, as it contains 82.5\% of Bi$_2$Te$_3$ pure phase and has  {desired} nanostructural morphology. So, we analyzed its seebeck coefficient, electrical conductivity, electrical resistivity and  {power factor(PF)} in between the temperature range of 310 K -398 K, using Seebeck Coefficient and Electrical Resistivity System (Ulvac ZEM-3) instrument.
 {Fig. \ref{fig8}(a) shows the electrical conductivity and resistivity curves with change in temperature with blue \& green curves, respectively, while Fig. \ref{fig8}(b) represents} change in seebeck coefficient, electrical conductivity \& power factor with temperature, represented by black, blue \& red curves, respectively.

For pure Bi$_2$Te$_3$, the electrical resistivity firstly increased from 100 K to 300 K, and then starts decreasing above 300 K\cite{a2}.The same decreasing pattern of resistivity has been observed from H5 sample between 310 K to 398 K. In this temperature range the value of electrical conductivity increases with the increase in temperature as electrical conductivity and resistivity are inversely proportional to each other. The electrical conductivity has a maximum value of 82.6 Sm$^{-1}$ at 400 K, which is very good as compared to the pure Bi$_2$Te$_3$, which have a conductivity of $\textless$4 Sm$^{-1}$ at 400 K\cite{C555}, as  {this will  results in higher ZT value}.
{Jeon \etal \cite{a2} suggested, for the pure Bi$_2$Te$_3$ system, such an increase in electrical conductivity attributes to the exponential increment of charge carrier concentration, as a result of onset of mixed conduction. Samanta \etal \cite{samanta} had done a comparative study on the thermoelectric properties of BiSe and Bi$_2$Se$_3$ systems, and observed that the value of electrical conductivity for BiSe is comparatively higher, as compared to Bi$_2$Se$_3$. This higher value of electrical conductivity attributes to the lowering in band gap of BiSe($\sim$20 meV) as compared to Bi$_2$Se$_3$ ($\sim$ 0.35 eV), as a result of additional Bi$_2$-layers present in between the quintuple layers[Se-Bi-Se-Bi-Se] in BiSe crystal structure as shown in Fig. \ref{fig1.5}(crystal structures of the compounds of (Bi$_2$)$_m$(Bi$_2$Se$_3$)$_n$ and (Bi$_2$)$_m$(Bi$_2$Te$_3$)$_n$ homologous series are almost same). {Similar case is}  {observed in case of H5 sample}, because of the extra Bi$_2$-layers present in between the Bi$_2$Te$_3$ quintuple in the phases like BiTe and Bi$_4$Te$_3$, which are present in smaller amount, the band gap might  {have} decreased, as both Te and Se belongs to the same chalcogenide family of periodic table and both Bi$_2$Se$_3$(0.35 eV) and Bi$_2$Te$_3$(0.15 eV) \cite{bg01} have comparable band gaps, although there are some certain differences in the band gaps structures of both Bi$_2$Se$_3$ and Bi$_2$Te$_3$ \cite{mishra}.}

 {From Fig. \ref{fig8}(b), the }negative value of seebeck coefficient shows that it is an $n$-type semiconductor. The value of seebeck coefficient for pure Bi$_2$Te$_3$ phase is nearly -220 $\mu$VK$^{-1}$ at 310 K \cite{C555}, but in our case, the value of seebeck coefficient is somewhat low (-68 $\mu$VK$^{-1}$ at 310 K) compared to the pure Bi$_2$Te$_3$ pure phase. This is due to the presence of impurity phases of homologous (Bi$_2$)$_m$(Bi$_2$Te$_3$)$_n$ series, like BiTe, Bi$_4$Te$_3$ and some unrefined phases present in the system. The value of seebeck coefficient initially decreases from -68 $\mu$VK$^{-1}$ at 310 K to -64.5 $\mu$VK$^{-1}$ between 350-373 K, where it becomes the maximum. After 373 K, there is {slight} increase in the seebeck coefficient value up to -64.7 $\mu$VK$^{-1}$ at 398 K.
{Because of the presence of lower band gap, the number of charge carriers in BiSe are more, which lowers the seebeck coefficient for BiSe, compared to Bi$_2$Se$_3$, but still the electronic density of state near the fermi level is highly asymmetrical for BiSe, compared to that of Bi$_2$Se$_3$, which gives BiSe a considerable value of seebeck coefficient despite of its higher number of charge carriers.  {The excess of Bi present in the Bi$_2$Te$_3$, adds the degeneracy region in the crystal, this will also decreased the Seebeck coefficient\cite{r1996}, as degeneracy implies the equalization of energy levels, while Seebeck coefficient is related to the difference in the fermi energy levels}. Since, our sample is a mixture of Bi$_2$Te$_3$, BiTe and other phase which have  {Bi$_2$-bilayer, s}o the value of $S$ is somewhat, intermediate between pure BiTe \& pure Bi$_2$Te$_3$ phase.}

The  {power factor(PF)} of a thermoelectric material is represented as
\begin{equation}
PF = S^2\sigma             \label{eq4.1}
\end{equation}
In order to achieve a higher $ZT$ value, the power factor of the material should be high. Power factor directly depends upon the seebeck coefficient and electrical conductivity. For H5 sample, change in power factor with an increase in temperature  {is given in Fig. \ref{fig8}(b).} Power factor of this sample initially remains constant$\sim$ 0.204 $\mu$W/m K$^2$ between 310-326 K, and then increases sharply.  {The maximum value of power factor(PF)} for the H5 sample, is 0.346 $\mu$W/m K$^2$, which is observed at 398 K.  {The pure phase of Bi$_2$T$_3$ has a power factor$\sim$ 1 mW/mK$^2$ at 400 K \cite{C555}, which  {is approximately 10$^3$ times greater} as compared to the H5 sample.} This lower value of power factor {is attributed} to the lower seebeck coefficient of the system.

\section{Conclusion and future perspective}
In summary, alternate methods are adopted for synthesis of pure phase of bismuth telluride as presence of impure phase deteriorate the thermoelectric performance of the material. In our hydrothermally synthesized sample case, we have observed that increasing the synthesis duration resulted in less impure Bi$_2$Te$_3$ phase along with change in morphology of the system. While increasing the synthesis temperature led to the decrease in size of the particles and small decrease in Bi$_2$Te$_3$ phase, as Te gets desorbed from the surface of the particles and results into the formation of phase which have lesser Te:Bi atomic ratio's, like BiTe and Bi$_4$Te$_3$ etc. The hydrothermal method is more rapid and efficient for preparation of bismuth telluride than co-precipitation methods. A more pure Bi$_2$Te$_3$ phase has been synthesized by preparing the sample through hydrothermal method,  {as compared with th}e co-precipitation method in which additional phases such as BiTe, Bi$_4$Te$_3$,TeO$_2$ were formed in greater amounts along with Bi$_2$Te$_3$.  {Focusing on the thermoelectric properties, the electrical conductivity of our best samples is {observed to be approximately} 20 times more than that of previously reported Bi$_2$Te$_3$ pure phase. This increase in electrical conductivity attributes to the lowering of bandgap, which is caused by additional Bi$_2$-layer in between the quintuple layers [Te-Bi-Te-Bi-Te] of additional BiTe and Bi$_4$T$_3$ phases. The seebeck coefficient is $\sim$3 times lower than that of pure Bi$_2$Te$_3$ phase because of higher number of charge carriers. Although, Bi$_2$-layers decreases the seebeck coefficient, but it also decreases the lattice thermal conductivity, through soft localized vibrations of this Bi$_2$-layers in the system, which is helpful in  {obtaining higher ZT values.\cite{samanta}}}

{ {Particle size} could be decreased by further increasing the synthesis temperature, along with optimizing the rate of evaporation of Te from the particle surfaces, so that, a more pure phase of Bi$_2$Te$_3$ can be obtained. Doping of suitable element in different phases of (Bi$_2$)$_m$(Bi$_2$Te$_3$)$_n$ series, which consists of Bi$_2$ bilayer, could result into higher seebeck coefficient \& can further improve the power factor of the system, as these already have higher electrical conductivity and lower lattice thermal conductivity, which can be helpful in obtaining a good value of ZT.}

\section*{Conflicts of interest}
All contributing authors declare no conflicts of interest.

\section*{Credit author statement}
{\bf S.Gautam :} conceptualization, methodology, supervision {\bf  {V. Thakur }\& K. Upadhyay:} data curation, writing - original draft preparation {\bf R. Kaur \& \bf V. Thakur:} synthesis \& experiments and draft writing {\bf N.Goyal:} supervision

\section{Acknowledgement}
This work was financially supported by TEQIP-III (World bank project, MHRD, Govt. of India). 
SG acknowledges the Department of Science and Technology, India (JNC/Synchrotron and Neutron/2019/IN-011. dated Oct 31, 2019) for the financial support and Jawaharlal Nehru Centre for Advanced Scientific Research (JNCASR) for managing the project. Mr. Krishan Sethi is acknowledged for continuous technical support in AFM Lab. {National Physical Laboratory (Dr. Bhasker Gahtori, NPLONE scheme) is acknowledged for the IV and seebeck measurements. I am thankful to all medical staff and researchers who are fighting against this COVID-19 pandemic, best wishes for their good health and deep condolence for all who lost their life in fighting against this pandemic.}
%
\bibliography{bite-bib}
\end{document}